\documentclass[12pt, fleqn]{article}

\usepackage{dsfont}

\usepackage{tikz}
\usetikzlibrary{calc}
\usetikzlibrary{decorations.pathreplacing}
\usetikzlibrary{decorations.markings}
\usetikzlibrary{intersections} 
\usetikzlibrary{er,positioning}
\usetikzlibrary{arrows,shapes}
\usetikzlibrary{decorations.markings}
\usetikzlibrary{decorations.pathreplacing,calligraphy}
\usepackage{BOONDOX-cal} 

\usepackage[title]{appendix}
\usepackage[margin=2cm]{geometry}

\usepackage{amsmath, amssymb, amsthm, amsfonts}

\usepackage[colorlinks=true, linktoc=all, linkcolor=black, citecolor=red, urlcolor=blue]{hyperref}

\usepackage{etoolbox}
\makeatletter
\patchcmd{\BR@backref}{\newblock}{\newblock(}{}{}
\patchcmd{\BR@backref}{\par}{)\par}{}{}
\makeatother

\numberwithin{equation}{section}
\usepackage{empheq}

\newcommand{\n}{\mathfrak{n}}

\newcommand{\TL}{\mathcal{T\!\!L}}

\newcommand{\uJTL}{u\mathcal{J\!T\!\!L}}
\newcommand{\dJTL}{d\mathcal{J\!T\!\!L}}

\newcommand{\dTL}{d\mathcal{T\!\!L}}

\newcommand{\DD}{\mathsf{D}}
\newcommand{\dd}{\mathsf{d}}
\newcommand{\II}{\mathds{1}}
\newcommand{\VirN}{\hbox{Vir}\otimes\overline{\hbox{Vir}}}

\begin{document}

\begin{flushleft}
{\bfseries\sffamily\Large 

Non-invertible symmetries  and RG flows 
 in the two-dimensional $O(n)$  loop model
\vspace{1.5cm}
\\
\hrule height .6mm
}
\vspace{1.5cm}

{\bfseries\sffamily 
Jesper Lykke Jacobsen$^{1,2,3}$,  
Hubert Saleur$^{1,4}$
}
\vspace{4mm}

{\textit{\noindent
$^1$ Institut de physique th\'eorique, CEA, CNRS, 
Universit\'e Paris-Saclay
\\
$^2$ Laboratoire de Physique de l'\'Ecole Normale Sup\'erieure, ENS, Universit\'e PSL, CNRS, Sorbonne Universit\'e, Universit\'e de Paris
\\ 
$^3$ Sorbonne Universit\'e, \'Ecole Normale Sup\'erieure, CNRS,
Laboratoire de Physique (LPENS)
\\ 
$^4$ Department of Physics and Astronomy, University of Southern California, Los Angeles
}}
\vspace{4mm}

\end{flushleft}
\vspace{7mm}

{\noindent\textsc{Abstract:}
In a recent paper, Gorbenko and Zan \cite{GZ} observed that $O(n)$ symmetry alone does not  protect  the well-known renormalization group flow from the dilute to the dense phase of the two-dimensional $O(n)$ model under thermal perturbations. We show in this paper that the required ``extra protection'' is topological in nature, and is related to the existence of certain non-invertible topological defect lines. We define these defect lines and discuss the ensuing topological protection, both in the context of the $O(n)$ lattice model and in its recently understood continuum limit, which takes the form of a conformal field theory governed by an interchiral algebra.
}

\clearpage

\hrule 
\tableofcontents
\vspace{5mm}
\hrule
\vspace{5mm}

\hypersetup{linkcolor=blue}

\section{Introduction}

In the conclusion to their paper, Gorbenko and Zan  \cite{GZ} raise a puzzling question about our understanding of the phase diagram of the $O(n)$ loop model. The puzzle occurs because, in the $O(n)$ model at the (dilute) critical point, geometrically well-identified fields with  given, known conformal dimensions, come with multiplicities that correspond to {\sl sums}  of several $O(n)$ irreducible representations. For instance, the four-leg (fuseau, or watermelon) operator comes with the multiplicity space $[4]\oplus [22]\oplus [211]\oplus [2]\oplus [\,]$, where we used the standard row-length notation $[\lambda]$ for (Young diagrams of) irreducible representations  of $O(n)$ (see (\ref{l20}) for the origin of this decomposition). This multiplicity means that there are copies of the four-leg operator coming with different $O(n)$ tensorial content\footnote{In most of this paper, we follow most of the literature and treat $O(n)$ as an ``ordinary'' symmetry, while in fact $n$ is not an integer but a real (or even complex) number. This point is further discussed in the conclusion.}.

While having representations with four boxes  (one per leg) in this content    is fully expected,  the emergence of the $[2]$ and the $[\,]$ representations is more surprising. As discussed in \cite{JRS22}, this is related with the fact (to be discussed in more detail) that lines cannot cross, and  is interpreted mathematically in terms of branching rules from the Brauer algebra down to  the (so-called ``unoriented Jones'') Temperley-Lieb algebra. What matters for the puzzle raised in \cite{GZ} is that there exists a copy of the four-leg operator  coming in the trivial representation---i.e.\ an {\sl $O(n)$ scalar}.

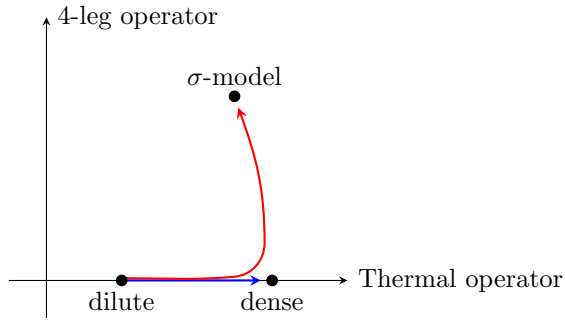
\begin{figure}
\centering
\begin{tikzpicture}[scale=1.0]
 \draw [->,>=stealth] (-0.5,0) -- (4,0);
 \draw [->,>=stealth] (0,-0.5) -- (0,3.5);
 \draw [thick,blue,->,>=stealth] (1,0)--(2.85,0);
 \draw [thick,red,->,>=stealth] (1,0.03) to [out=0,in=185] (2.5,0.05) to [out=5,in=-90] (2.9,0.5) to [out=90,in=-70] (2.55,2.3);
 \draw (0,3.5) node[right] {\footnotesize 4-leg operator};
 \draw (4,0) node[right] {\footnotesize Thermal operator};
 \draw [black,fill] (1,0) circle(2pt);
 \draw (1,0) node[below] {\footnotesize dilute};
 \draw [black,fill] (3,0) circle(2pt);
 \draw (3,0) node[below] {\footnotesize dense};
 \draw [black,fill] (2.5,2.45) circle(2pt);
 \draw (2.5,2.45) node[above] {\footnotesize $\sigma$-model};
\end{tikzpicture}
\caption{With only $O(n)$ symmetry  ``protection'', the 4-leg operator shold be generated under the RG, and the flow go to the $\sigma$-model fixed point \cite{JReadS} instead of the dense one.}
\label{Fig0}
\end{figure}

Now imagine perturbing the critical $O(n)$ model by the energy operator, itself an $O(n)$ scalar with conformal weights $\Delta=\bar{\Delta}=\Delta_{(1,3)}$ (see below for conventions). This perturbation is, for the sign that corresponds to increasing the fugacity of monomers, known to drive the model into the dense critical phase \cite{Nienhuis,DuplantierSaleur}. With respect to this renormalization group (RG) flow, the four-leg operator is {\sl dangerously irrelevant}: while it is irrelevant at the (dilute) ultraviolet (UV) critical point, it is relevant at the dense infrared (IR) critical point. For instance, the case of self-avoiding walks (SAW) corresponds to the $n \to 0$ limit, where the corresponding scaling dimensions $x = \Delta + \bar{\Delta}$ at either end of the RG flow take the values $x_{\rm UV}={35\over 12}>2$ and $x_{\rm IR}={3\over 4}<2$. Since there is a copy of the four-leg operator that comes with the scalar $O(n)$ representation, the $O(n)$ symmetry by itself is not sufficient to prevent this operator from ``interfering"  with the RG flow induced by the energy-operator. With only $O(n)$ symmetry, there is therefore no reason why the  dense fixed point should ever be reached: rather, one should expect an RG  trajectory where models get close enough to that fixed point, but eventually reach the more generic $O(n)$ low-temperature phase studied in \cite{JReadS,NSSO13,Granet19}. See the cartoon in figure \ref{Fig0}.

\smallskip

Yet, there is large numerical evidence over more than 30 years of research that this is not what happens, and that in all cases (in particular,  irrespectively of the lattice) the dense fixed point is indeed reached by the RG flow when the monomer fugacity is increased beyond the critical value, without undue interference from the dangerously irrelevant four-leg operator. {\sl This is the puzzle}.

Of course, the fact that the four-leg operator comes with a sum of $O(n)$ representations suggests that there is a hidden symmetry at play---a symmetry glueing together these various representations, and for which the  copy of the trivial   $O(n)$ representation $[\,]$ appearing in the multiplicity of the four-leg operator  would not be a scalar anymore. It is not clear, however, what this hidden symmetry might be: attempts to find the corresponding extra generators are difficult, and there is evidence that it is not possible to define the necessary associated tensor products. 

Reference \cite{GZ} makes two final suggestions to solve the puzzle: one involves integrability, the other the presence of a yet unidentified symmetry involving {\sl non-invertible topological defect lines}. We shall now demonstrate that the latter suggestion is correct, and analyze it in detail.

\bigskip

The paper is organized as follows. The set-up and the general idea how to introduce  the relevant topological defect lines in the loop model are given in section 2. Section 3 provides technical details of the construction for the lattice model, in the dense and dilute cases. Section 4 discusses the ensuing properties at the fixed points and along the RG flow. This concludes the goal of this paper proper. We find it useful, however, to discuss briefly further aspects in section 5, and to establish in particular a connection with the so-called Verlinde lines \cite{CSLWY}. Obviously, our work opens the way to considering many other questions in loop models, some of which are mentioned in the conclusion. 

The paper also contains three appendices. In appendix A we discuss the claim that crossings do not change the universality class at the dilute fixed point. In appendix B we consider what happens to our topological defect lines with open boundary conditions and  in appendix C, what happens when one tries to build similar objects in models with crossings.

\section{General idea}

Topological defects  are conveniently introduced  in CFTs (see e.g.\ \cite{PetkovaZuber,Juergen} for early references) as operators%
\footnote{The name ``defect operator'' per se  is used for a different object in  the literature \cite{CSLWY}.} 
acting on the Hilbert space ${\cal H}$ of the theory, which it itself a sum of representations of $\VirN$. These operators, which we will denote generically by $\DD$, must commute with $\hbox{Vir}$ and $\overline{\hbox{Vir}}$:
\begin{eqnarray}
\left[\DD,\mathsf{L}_p\right]=\left[\DD,\bar{\mathsf{L}}_p\right]=0 \,, \quad \forall p\in \mathbb{Z}\label{commdef}
\end{eqnarray}
or, equivalently, commute independently with the left- and right-moving components of the stress-energy tensor. 

The corresponding objects in the Euclidian description are  topological defect-lines (TDL), which we denote by the  symbol $D$. When correlation functions are calculated in their presence, relations (\ref{commdef})  mean that the lines can be deformed at will without changing the result, provided the  field insertions are not crossed. This latter definition extends easily to the non-conformal case. For a more thorough set of definitions, see e.g.\ \cite{CSLWY}.

While these definitions are well known in the continuum limit, they can be extended to lattice models (see e.g.\ \cite{AMF} for a discussion in the context of the Ising model). Radial quantization corresponds then to studying the model on a cylinder with periodic boundary conditions in the space direction, while the Hamiltonian (or, more generally, the transfer matrix) propagates in the imaginary-time direction. It is natural to define a lattice equivalent of $\DD$  in this context as an operator that commutes with the local densities or energy and momentum---this is discussed in more detail below. In the Euclidian description, the lattice defect line should be---just like in the continuum---deformable at will without changing correlation functions, provided it does not cross insertions of discrete lattice observables. 

Since we are interested in systems with global $O(n)$ symmetry---and forgetting for now the fact that $n$ is in general  non-integer  in this paper---so-called invertible  topological lines (i.e.\ associated with the action of a group element $g$)  can easily be constructed, and were in fact part of the analysis in \cite{JRS22}. In the present paper we are however concerned with a different type
of  lines which are {\sl non-invertible} and not related to a group element in any obvious way.
We now claim that there exists a natural pair of such lines  in the lattice $O(n)$ model answering the suggestion in \cite{GZ}: the  presence of the associated operators $\DD,\overline{\DD}$  (or their lattice equivalent) will be called  ``non-invertible (topological) symmetry''. Before dwelling into technicalities, let us give a qualitative idea of how this works.

We start by recalling that the partition function of the $O(n)$ model is given by an expression of the type
\begin{equation}
Z_{O(n)}=\sum_{\mathcal {C}} 
K^{|A|(\mathcal{C})} n^{N(\mathcal{C})} \,, \label{Zexpan}
\end{equation}
where the sum is over self- and mutually-avoiding loop configurations  $\mathcal{C}$ (no crossings, no overlaps), $N(\mathcal{C})$ is the number of loops so $n$ is interpreted as their fugacity, $|A|(\mathcal{C})$ is the number of lattice edges (``monomers'')  involved in $\mathcal{C}$, and $K$ is their fugacity. The $O(n)$ (dilute) fixed point is obtained for a particular critical value, $K=K_{\rm c}$; the model is then in  the dense critical phase  and flows to the dense critical fixed point (whose properties are independent of $K$) for all $K>K_{\rm c}$. We will consider more generally a model where the monomer couplings are edge-dependent, with a generalized partition function
\begin{equation}
Z_{O(n)}=\sum_{\mathcal {C}}\left( \prod_{e \in A(\mathcal{C})}K_e\right) n^{N(\mathcal{C})} \,, \label{Zexpan1}
\end{equation}
where now the product is taken over the set of monomers $A(\mathcal{C})$ in each configuration $\mathcal{C}$. The best known observables in this model are the fuseau operators, sources and sinks of  $2r$ lines which cannot be contracted among themselves, but which can only connect to another fuseau operator.

While the definition (\ref{Zexpan}) is purely geometrical, we also recall that it arises from a high-temperature expansion for a model  of spins living on the edges or the vertices of the lattice (depending on the particular
underlying microscopic model). These spins belong to the vector (fundamental) representation of $O(n)$, denoted $[1]$. This is in particular what explains the fugacity of loops---the factor $n$ corresponds to the $n$ colors that can ``propagate'' around it. The role of the $O(n)$ symmetry will become much clearer from the further details given below. It is convenient in what follows to imagine that we have some sort of continuum version of (\ref{Zexpan})--(\ref{Zexpan1}), where $K$ is coupled to a properly defined ``length'' of the loops, and to forget for the time being about the lattice.

We now introduce into this system another type of geometrical object, representing a {\sl topological defect line}. This line (which can be in particular  a closed loop) goes ``above'' the loops of the $O(n)$ model. To give a precise meaning to this, we parameterize
\begin{equation}
 n = q + q^{-1} \,,
\end{equation}
and we interpret over- and under-crossings as giving rise to the decompositions
\begin{subequations}
\label{Fig1}
\begin{eqnarray}
\begin{tikzpicture}[scale=0.8,baseline={([yshift=-3pt]current bounding box.center)}]
 \draw [thick,red] (1,0)--(0,1);
 \draw [thick,blue] (0,0) -- (0.45,0.45);
 \draw [thick,blue] (0.55,0.55) -- (1,1);
\end{tikzpicture} &=& (-q)^{1/2}
\begin{tikzpicture}[scale=0.8,baseline={([yshift=-3pt]current bounding box.center)}]
 \draw [thick,blue] (0,0) arc(-45:0:1.5 and 0.707);
 \draw [thick,red] (0,1) arc(45:0:1.5 and 0.707);
 \draw [thick,red] (1,0) arc(-135:-180:1.5 and 0.707);
 \draw [thick,blue] (1,1) arc(135:180:1.5 and 0.707);
\end{tikzpicture} + (-q)^{-1/2}
\begin{tikzpicture}[scale=0.8,baseline={([yshift=-3pt]current bounding box.center)}]
 \draw [thick,blue] (0,0) arc(135:90:0.707 and 1.5);
 \draw [thick,red] (0,1) arc(-135:-90:0.707 and 1.5);
 \draw [thick,red] (1,0) arc(45:90:0.707 and 1.5);
 \draw [thick,blue] (1,1) arc(-45:-90:0.707 and 1.5);
\end{tikzpicture} \,, \\
\begin{tikzpicture}[scale=0.8,baseline={([yshift=-3pt]current bounding box.center)}]
 \draw [thick,red] (0,0)--(1,1);
 \draw [thick,blue] (0,1) -- (0.45,0.55);
 \draw [thick,blue] (0.55,0.45) -- (1,0);
\end{tikzpicture} &=& (-q)^{1/2}
\begin{tikzpicture}[scale=0.8,baseline={([yshift=-3pt]current bounding box.center)}]
 \draw [thick,red] (0,0) arc(135:90:0.707 and 1.5);
 \draw [thick,blue] (0,1) arc(-135:-90:0.707 and 1.5);
 \draw [thick,blue] (1,0) arc(45:90:0.707 and 1.5);
 \draw [thick,red] (1,1) arc(-45:-90:0.707 and 1.5);
\end{tikzpicture} + (-q)^{-1/2}
\begin{tikzpicture}[scale=0.8,baseline={([yshift=-3pt]current bounding box.center)}]
 \draw [thick,red] (0,0) arc(-45:0:1.5 and 0.707);
 \draw [thick,blue] (0,1) arc(45:0:1.5 and 0.707);
 \draw [thick,blue] (1,0) arc(-135:-180:1.5 and 0.707);
 \draw [thick,red] (1,1) arc(135:180:1.5 and 0.707);
\end{tikzpicture} \,,
\end{eqnarray} 
\end{subequations}
where $O(n)$ lines are represented in blue and topological lines in red. After the expansion we obtain lines or loops with two colors. The colors do not imply any distinction by themselves---after all, loops of any color live in the same representation [1]---but they provide a nice visual means of tracking the role of the topological lines in the decompositions. The rule is that we give to every loop, irrespective of its (bi)color the weight $n$, while the variable $K$ is only coupled to the length of the blue parts,%
\footnote{This is only a question of convention: since the topological defect line is not a fluctuating object, giving its monomers a non-trivial  weight can easily be taken  into account by adjusting normalizations.}
which is not affected by the  topological line.  

The choice of the coefficients guarantees that the topological line(s) can be moved at will without changing partition or correlation functions, justifying the graphical representation that hints at invariance under topological (Reidemeister) moves. A short calculation to illustrate this point goes as follows:
\begin{eqnarray}
\label{Fig2}
\begin{tikzpicture}[scale=0.69,baseline={([yshift=-3pt]current bounding box.center)},rotate=-45]
 \draw [thick,blue] (0,0) arc(-90:-5:0.5 and 1);
 \draw [thick,blue] (0,0) arc(-90:-175:0.5 and 1);
 \draw [thick,blue] (0,2) arc(90:175:0.5 and 1);
 \draw [thick,blue] (0,2) arc(90:5:0.5 and 1);
 \draw [thick,red] (0,2.2) arc(90:450:1.2 and 0.63);
\end{tikzpicture} &=& (-q)
\begin{tikzpicture}[scale=0.69,baseline={([yshift=-3pt]current bounding box.center)},rotate=-45]
 \draw [thick,red] (-0.5,1) to [out=135,in=-90] (-1.2,1.6) to [out=90,in=180] (0,2.2) to [out=0,in=90] (1.2,1.6) to [out=-90,in=-45] (0.5,1.1);
 \draw [thick,blue] (0.5,1.1) to [out=135,in=0] (0,2) to [out=180,in=135] (-0.5,1.1);
 \draw [thick,blue] (0.5,1) to [out=-45,in=0] (0,0) to [out=180,in=-45] (-0.5,1);
 \draw [thick,red] (-0.5,1.1) to [out=-45,in=135] (0.5,1);
\end{tikzpicture} +
\begin{tikzpicture}[scale=0.69,baseline={([yshift=-3pt]current bounding box.center)},rotate=-45]
 \draw [thick,red] (-0.5,1) to [out=135,in=-90] (-1.2,1.6) to [out=90,in=180] (0,2.2) to [out=0,in=90] (1.2,1.6) to [out=-90,in=45] (0.5,1);
 \draw [thick,blue] (0.5,1) to [out=225,in=0] (0,0) to [out=180,in=-45] (-0.5,1);
 \draw [thick,red] (-0.44,1.1) to [out=-90,in=270] (0.44,1.1);
 \draw [thick,blue] (-0.44,1.1) to [out=90,in=180] (0,2) to [out=0,in=90] (0.44,1.1);
\end{tikzpicture} +
\begin{tikzpicture}[scale=0.69,baseline={([yshift=-3pt]current bounding box.center)},rotate=-45]
 \draw [thick,red] (-0.5,1.1) to [out=225,in=-90] (-1.2,1.6) to [out=90,in=180] (0,2.2) to [out=0,in=90] (1.2,1.6) to [out=-90,in=-45] (0.5,1.1);
 \draw [thick,blue] (-0.5,1.1) to [out=45,in=180] (0,2) to [out=0,in=135] (0.5,1.1);
 \draw [thick,blue] (0.5,1) to [out=315,in=0] (0,0) to [out=180,in=-135] (-0.5,1);
 \draw [thick,red] (-0.5,1) to [out=45,in=135] (0.5,1);
\end{tikzpicture} + (-q)^{-1}
\begin{tikzpicture}[scale=0.69,baseline={([yshift=-3pt]current bounding box.center)},rotate=-45]
 \draw [thick,red] (-0.5,1.1) to [out=225,in=-90] (-1.2,1.6) to [out=90,in=180] (0,2.2) to [out=0,in=90] (1.2,1.6) to [out=-90,in=45] (0.5,1.0);
 \draw [thick,blue] (0.5,1.1) to [out=45,in=0] (0,2) to [out=180,in=45] (-0.5,1.1);
 \draw [thick,blue] (0.5,1) to [out=-135,in=0] (0,0) to [out=180,in=-135] (-0.5,1);
 \draw [thick,red] (-0.5,1.0) to [out=45,in=-135] (0.5,1.1);
\end{tikzpicture} \nonumber \\ &=& (q+q^{-1})
\begin{tikzpicture}[scale=0.69,baseline={([yshift=-3pt]current bounding box.center)},rotate=-45]
 \draw [thick,blue] (0,0) arc(-90:0:0.5 and 1);
 \draw [thick,blue] (0,0) arc(-90:-180:0.5 and 1);
 \draw [thick,blue] (0,2) arc(90:180:0.5 and 1);
 \draw [thick,blue] (0,2) arc(90:0:0.5 and 1);
\end{tikzpicture}
\end{eqnarray}
The final result means that red loops can be moved around without affecting the partition function,
as summarized in the graphical identity
\begin{equation}
\label{Fig3}
\begin{tikzpicture}[scale=0.69,baseline={([yshift=-3pt]current bounding box.center)},rotate=-45]
 \draw [thick,blue] (0,0) arc(-90:-5:0.5 and 1);
 \draw [thick,blue] (0,0) arc(-90:-175:0.5 and 1);
 \draw [thick,blue] (0,2) arc(90:175:0.5 and 1);
 \draw [thick,blue] (0,2) arc(90:5:0.5 and 1);
 \draw [thick,red] (0,2.2) arc(90:450:1.2 and 0.63);
\end{tikzpicture} =
\begin{tikzpicture}[scale=0.69,baseline={([yshift=-3pt]current bounding box.center)},rotate=-45]
 \draw [thick,blue] (0,0) arc(-90:270:0.5 and 1);
 \draw [thick,red] (1.5,3.0) arc(90:450:1.2 and 0.63);
\end{tikzpicture}
\end{equation}

The possibility of moving the TDL  around arises---as discussed below---from the fact that the combination on the right-hand side of equation (\ref{Fig1}) is a solution of the spectral-parameter independent Yang-Baxter equation. 
Without field insertions, adding a topological loop simply multiplies the partition function by a factor $n$. We shall denote in what follows statistical averages associated with the partition function  by $\langle \cdot \rangle$, leading to 
\begin{equation}
\langle {D}\rangle=n \,.\label{gsexpect}
\end{equation}
We can of course write a similar statement on the cylinder. Denoting the (normalized) ground state of the model by $|\II\rangle$ we have then
\begin{equation}
\langle \II|\DD|\II\rangle=n \,.
\end{equation}
 It is of course possible to have the defect line go under instead of over the loops. The rules will be the same as before (see figure \ref{Fig1}), but with red and blue colors swapped. We will refer to the corresponding objects as $\overline{\DD},\overline{D}$. 

While seemingly trivial when calculating  the partition function, this TDL is non-trivial when inserted into correlation functions. To illustrate this point we may consider the two-point function of the one-leg operator in the presence of a TDL. This gives rise to the calculation%
\footnote{The potential ambiguity in the determination of the square root goes back to \eqref{Fig1}. The reason that we do not factor the square root out of \eqref{Fig1} is that those definitions ensure the Reidemeister move II when pulling the strings in any direction.}

\begin{equation}
\label{Fig5}
\begin{tikzpicture}[scale=0.69,baseline={([yshift=-3pt]current bounding box.center)},rotate=-45]
 \draw [thick,red] (0,0) arc(0:360:1 and 1);
 \draw [thick,blue] (-1,0) -- (-1,0.9);
 \draw [thick,blue] (-1,1.1) -- (-1,2);
 \draw [thick,black] (-1.1,-0.1) -- (-0.9,0.1);
 \draw [thick,black] (-1.1,0.1) -- (-0.9,-0.1);
 \draw [thick,black] (-1.1,1.9) -- (-0.9,2.1);
 \draw [thick,black] (-1.1,2.1) -- (-0.9,1.9);
\end{tikzpicture} = (-q)^{1/2}
\begin{tikzpicture}[scale=0.69,baseline={([yshift=-3pt]current bounding box.center)},rotate=-45]
 \draw [thick,red]  (-1,0.9) to [out=135,in=90] (-2,0) to [out=-90,in=-180] (-1,-1) to [out=0,in=-90] (0,0) to [out=90,in=-45] (-1,1.1);
 \draw [thick,blue] (-1,0) to [out=90,in=-45] (-1,0.9);
 \draw [thick,blue] (-1,1.1) to [out=135,in=-90] (-1,2);
 \draw [thick,black] (-1.1,-0.1) -- (-0.9,0.1);
 \draw [thick,black] (-1.1,0.1) -- (-0.9,-0.1);
 \draw [thick,black] (-1.1,1.9) -- (-0.9,2.1);
 \draw [thick,black] (-1.1,2.1) -- (-0.9,1.9);
\end{tikzpicture} + (-q)^{-1/2}
\begin{tikzpicture}[scale=0.69,baseline={([yshift=-3pt]current bounding box.center)},rotate=-45]
 \draw [thick,red]  (-1,1.1) to [out=-135,in=90] (-2,0) to [out=-90,in=-180] (-1,-1) to [out=0,in=-90] (0,0) to [out=90,in=45] (-1,0.9);
 \draw [thick,blue] (-1,0) to [out=90,in=-135] (-1,0.9);
 \draw [thick,blue] (-1,1.1) to [out=45,in=-90] (-1,2);
 \draw [thick,black] (-1.1,-0.1) -- (-0.9,0.1);
 \draw [thick,black] (-1.1,0.1) -- (-0.9,-0.1);
 \draw [thick,black] (-1.1,1.9) -- (-0.9,2.1);
 \draw [thick,black] (-1.1,2.1) -- (-0.9,1.9);
\end{tikzpicture}
\end{equation}
On the cylinder, denoting by $|(r,0)\rangle$ the ground state of the sector with $2r$ non-contractible lines---the notation is due to the operator inserting these lines having the Kac labels $(r,0)$---we will show later that
\begin{equation}
\langle (r,0)|\DD|(r,0)\rangle\equiv \DD_{(r,0)}=(-q)^{r}+(-q)^{-r} \,. \label{topeigen}
\end{equation}
Note that for $q$ generic, these numbers are different for different values or $r$, showing that the associated fields belong to different topological sectors.

In a nutshell, the rest of the argument will be that the thermal perturbation of the dilute fixed point corresponding to taking $K>K_{\rm c}$ preserves the non-invertible symmetry. Since the four-leg operator  is not the in the topological sector of the identity, it cannot be generated under RG. This explains the ``protection'' enjoyed by the model, and the possibility of having a flow to the dense fixed point in the IR. 

We now get into technicalities to put these ideas on firmer ground. In particular, while  we have been a bit cavalier in the foregoing discussion about distinguishing  the lattice and the continuum properties, we shall soon  see that the non-invertible symmetry can be properly defined for  the lattice model itself, whether critical or not. 

\section{Non-invertible (topological) symmetry in the lattice model}

\subsection{Algebras and the defect}

There are different lattice versions of the $O(n)$ model. While we believe their qualitative features and phase diagrams are universal (if crossings are not allowed), although some details might differ. We will consider here a version where the monomers live on the edges of the square lattice $S$. The interactions at one vertex of $S$ can be represented by drawing the possible states of a tile containing the given vertex and its adjacent half-edges:
\begin{equation}
\label{Fig6}
\begin{tikzpicture}[scale=0.85,baseline={([yshift=-3pt]current bounding box.center)}]
 \draw [gray] (-0.5,0)--(0,-0.5)--(0.5,0)--(0,0.5)--cycle;
 \draw [gray,fill] (-0.25,0.25) circle(1pt);
 \draw [gray,fill] (-0.25,-0.25) circle(1pt);
 \draw [gray,fill] (0.25,0.25) circle(1pt);
 \draw [gray,fill] (0.25,-0.25) circle(1pt);
\end{tikzpicture} \quad
\begin{tikzpicture}[scale=0.85,baseline={([yshift=-3pt]current bounding box.center)}]
 \draw [thick,blue] (-0.25,-0.25) to [out=45,in=-90] (-0.05,0) to [out=90,in=-45] (-0.25,0.25);
 \draw [gray] (-0.5,0)--(0,-0.5)--(0.5,0)--(0,0.5)--cycle;
 \draw [gray,fill] (-0.25,0.25) circle(1pt);
 \draw [gray,fill] (-0.25,-0.25) circle(1pt);
 \draw [gray,fill] (0.25,0.25) circle(1pt);
 \draw [gray,fill] (0.25,-0.25) circle(1pt);
\end{tikzpicture} \quad
\begin{tikzpicture}[scale=0.85,baseline={([yshift=-3pt]current bounding box.center)}]
 \draw [thick,blue] (0.25,-0.25) to [out=135,in=-90] (0.05,0) to [out=90,in=-135] (0.25,0.25);
 \draw [gray] (-0.5,0)--(0,-0.5)--(0.5,0)--(0,0.5)--cycle;
 \draw [gray,fill] (-0.25,0.25) circle(1pt);
 \draw [gray,fill] (-0.25,-0.25) circle(1pt);
 \draw [gray,fill] (0.25,0.25) circle(1pt);
 \draw [gray,fill] (0.25,-0.25) circle(1pt);
\end{tikzpicture} \quad
\begin{tikzpicture}[scale=0.85,baseline={([yshift=-3pt]current bounding box.center)}]
 \draw [thick,blue] (-0.25,-0.25) to [out=45,in=-180] (0,-0.05) to [out=0,in=135] (0.25,-0.25);
 \draw [gray] (-0.5,0)--(0,-0.5)--(0.5,0)--(0,0.5)--cycle;
 \draw [gray,fill] (-0.25,0.25) circle(1pt);
 \draw [gray,fill] (-0.25,-0.25) circle(1pt);
 \draw [gray,fill] (0.25,0.25) circle(1pt);
 \draw [gray,fill] (0.25,-0.25) circle(1pt);
\end{tikzpicture} \quad
\begin{tikzpicture}[scale=0.85,baseline={([yshift=-3pt]current bounding box.center)}]
 \draw [thick,blue] (-0.25,0.25) to [out=-45,in=180] (0,0.05) to [out=0,in=45] (0.25,0.25);
 \draw [gray] (-0.5,0)--(0,-0.5)--(0.5,0)--(0,0.5)--cycle;
 \draw [gray,fill] (-0.25,0.25) circle(1pt);
 \draw [gray,fill] (-0.25,-0.25) circle(1pt);
 \draw [gray,fill] (0.25,0.25) circle(1pt);
 \draw [gray,fill] (0.25,-0.25) circle(1pt);
\end{tikzpicture} \quad
\begin{tikzpicture}[scale=0.85,baseline={([yshift=-3pt]current bounding box.center)}]
 \draw [thick,blue] (-0.25,0.25) -- (0.25,-0.25);
 \draw [gray] (-0.5,0)--(0,-0.5)--(0.5,0)--(0,0.5)--cycle;
 \draw [gray,fill] (-0.25,0.25) circle(1pt);
 \draw [gray,fill] (-0.25,-0.25) circle(1pt);
 \draw [gray,fill] (0.25,0.25) circle(1pt);
 \draw [gray,fill] (0.25,-0.25) circle(1pt);
\end{tikzpicture} \quad
\begin{tikzpicture}[scale=0.85,baseline={([yshift=-3pt]current bounding box.center)}]
 \draw [thick,blue] (-0.25,-0.25) -- (0.25,0.25);
 \draw [gray] (-0.5,0)--(0,-0.5)--(0.5,0)--(0,0.5)--cycle;
 \draw [gray,fill] (-0.25,0.25) circle(1pt);
 \draw [gray,fill] (-0.25,-0.25) circle(1pt);
 \draw [gray,fill] (0.25,0.25) circle(1pt);
 \draw [gray,fill] (0.25,-0.25) circle(1pt);
\end{tikzpicture} \quad
\begin{tikzpicture}[scale=0.85,baseline={([yshift=-3pt]current bounding box.center)}]
 \draw [thick,blue] (-0.25,-0.25) to [out=45,in=-90] (-0.05,0) to [out=90,in=-45] (-0.25,0.25);
 \draw [thick,blue] (0.25,-0.25) to [out=135,in=-90] (0.05,0) to [out=90,in=-135] (0.25,0.25);
 \draw [gray] (-0.5,0)--(0,-0.5)--(0.5,0)--(0,0.5)--cycle;
 \draw [gray,fill] (-0.25,0.25) circle(1pt);
 \draw [gray,fill] (-0.25,-0.25) circle(1pt);
 \draw [gray,fill] (0.25,0.25) circle(1pt);
 \draw [gray,fill] (0.25,-0.25) circle(1pt);
\end{tikzpicture} \quad
\begin{tikzpicture}[scale=0.85,baseline={([yshift=-3pt]current bounding box.center)}]
 \draw [thick,blue] (-0.25,-0.25) to [out=45,in=-180] (0,-0.05) to [out=0,in=135] (0.25,-0.25);
 \draw [thick,blue] (-0.25,0.25) to [out=-45,in=180] (0,0.05) to [out=0,in=45] (0.25,0.25);
 \draw [gray] (-0.5,0)--(0,-0.5)--(0.5,0)--(0,0.5)--cycle;
 \draw [gray,fill] (-0.25,0.25) circle(1pt);
 \draw [gray,fill] (-0.25,-0.25) circle(1pt);
 \draw [gray,fill] (0.25,0.25) circle(1pt);
 \draw [gray,fill] (0.25,-0.25) circle(1pt);
\end{tikzpicture}
\end{equation}
For clarity we have drawn only the monomers (in blue color), omitting the lattice edges themselves.
This model results from the usual high-temperature expansion of a model with $O(n)$ spins living at the mid-points of the edges of $S$, represented as grey dots in (\ref{Fig6}), where we have also represented (still in grey color) the edges of the dual lattice $S^*$ that form the boundaries of the tile. A straight line or a corner 
on $S$ is counted as one monomer, so the first vertex  has no monomers, the  next six vertices  have one, and the last two vertices  have two. 

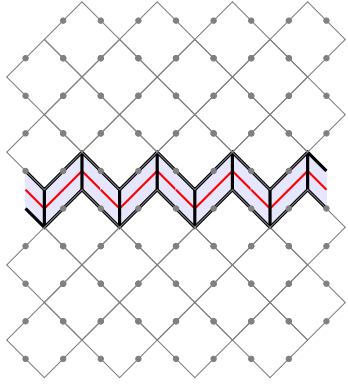
\begin{figure}
\centering
\begin{tikzpicture}[scale=1]
 \foreach \xpos in {1,2,3,4}{
 \draw [fill,blue!10] (-0.25+\xpos,1.75) -- (0+\xpos,1.5) -- (0.5+\xpos,2) -- (0.75+\xpos,1.75) -- (0.75+\xpos,2.25) -- (0.5+\xpos,2.5) --  (0+\xpos,2) -- (-0.25+\xpos,2.25) -- cycle; 
 \draw [thick,red] (-0.25+\xpos,2) -- (0+\xpos,1.75) -- (0.5+\xpos,2.25) -- (0.75+\xpos,2);
 \draw [very thick,black] (-0.25+\xpos,1.75) -- (0+\xpos,1.5) -- (0.25+\xpos,1.75);
 \draw [very thick,black] (-0.25+\xpos,2.25) -- (0+\xpos,2) -- (0.25+\xpos,2.25);
 \draw [very thick,black] (0+\xpos,1.5) -- (0+\xpos,2);
 \draw [very thick,black] (0.25+\xpos,1.75) -- (0.5+\xpos,2) -- (0.75+\xpos,1.75);
 \draw [very thick,black] (0.25+\xpos,2.25) -- (0.5+\xpos,2.5) -- (0.75+\xpos,2.25);
 \draw [very thick,black] (0.5+\xpos,2) -- (0.5+\xpos,2.5);
 }
 \foreach \ypos in {0,1,2.5,3.5}{
 \foreach \xpos in {1,2,3,4}{
 \draw [gray] (-0.5+\xpos,0+\ypos)--(0+\xpos,-0.5+\ypos)--(0.5+\xpos,0+\ypos)--(0+\xpos,0.5+\ypos)--cycle;
 \draw [gray,fill] (-0.25+\xpos,0.25+\ypos) circle(1pt);
 \draw [gray,fill] (-0.25+\xpos,-0.25+\ypos) circle(1pt);
 \draw [gray,fill] (0.25+\xpos,0.25+\ypos) circle(1pt);
 \draw [gray,fill] (0.25+\xpos,-0.25+\ypos) circle(1pt);
 \begin{scope}[xshift=0.5cm,yshift=0.5cm]
 \draw [gray] (-0.5+\xpos,0+\ypos)--(0+\xpos,-0.5+\ypos)--(0.5+\xpos,0+\ypos)--(0+\xpos,0.5+\ypos)--cycle;
 \draw [gray,fill] (-0.25+\xpos,0.25+\ypos) circle(1pt);
 \draw [gray,fill] (-0.25+\xpos,-0.25+\ypos) circle(1pt);
 \draw [gray,fill] (0.25+\xpos,0.25+\ypos) circle(1pt);
 \draw [gray,fill] (0.25+\xpos,-0.25+\ypos) circle(1pt);
 \end{scope}}}
\end{tikzpicture}
\caption{The defect line is introduced.}
\label{Fig7}
\end{figure}

It is convenient in what follows  to consider a ``natural propagation'' for the loops, corresponding to a diagonal propagation for the initial $O(n)$ spins---as in figure \ref{Fig7}. The transfer matrix
\begin{equation}
\label{Tmat}
T =
\begin{tikzpicture}[scale=0.85,baseline={([yshift=-3pt]current bounding box.center)}]
 \draw [gray] (-0.5,0)--(0,-0.5)--(0.5,0)--(0,0.5)--cycle;
 \draw [gray,fill] (-0.25,0.25) circle(1pt);
 \draw [gray,fill] (-0.25,-0.25) circle(1pt);
 \draw [gray,fill] (0.25,0.25) circle(1pt);
 \draw [gray,fill] (0.25,-0.25) circle(1pt);
\end{tikzpicture} + K \left(
\begin{tikzpicture}[scale=0.85,baseline={([yshift=-3pt]current bounding box.center)}]
 \draw [thick,blue] (-0.25,-0.25) to [out=45,in=-90] (-0.05,0) to [out=90,in=-45] (-0.25,0.25);
 \draw [gray] (-0.5,0)--(0,-0.5)--(0.5,0)--(0,0.5)--cycle;
 \draw [gray,fill] (-0.25,0.25) circle(1pt);
 \draw [gray,fill] (-0.25,-0.25) circle(1pt);
 \draw [gray,fill] (0.25,0.25) circle(1pt);
 \draw [gray,fill] (0.25,-0.25) circle(1pt);
\end{tikzpicture} +
\begin{tikzpicture}[scale=0.85,baseline={([yshift=-3pt]current bounding box.center)}]
 \draw [thick,blue] (0.25,-0.25) to [out=135,in=-90] (0.05,0) to [out=90,in=-135] (0.25,0.25);
 \draw [gray] (-0.5,0)--(0,-0.5)--(0.5,0)--(0,0.5)--cycle;
 \draw [gray,fill] (-0.25,0.25) circle(1pt);
 \draw [gray,fill] (-0.25,-0.25) circle(1pt);
 \draw [gray,fill] (0.25,0.25) circle(1pt);
 \draw [gray,fill] (0.25,-0.25) circle(1pt);
\end{tikzpicture} +
\begin{tikzpicture}[scale=0.85,baseline={([yshift=-3pt]current bounding box.center)}]
 \draw [thick,blue] (-0.25,-0.25) to [out=45,in=-180] (0,-0.05) to [out=0,in=135] (0.25,-0.25);
 \draw [gray] (-0.5,0)--(0,-0.5)--(0.5,0)--(0,0.5)--cycle;
 \draw [gray,fill] (-0.25,0.25) circle(1pt);
 \draw [gray,fill] (-0.25,-0.25) circle(1pt);
 \draw [gray,fill] (0.25,0.25) circle(1pt);
 \draw [gray,fill] (0.25,-0.25) circle(1pt);
\end{tikzpicture} +
\begin{tikzpicture}[scale=0.85,baseline={([yshift=-3pt]current bounding box.center)}]
 \draw [thick,blue] (-0.25,0.25) to [out=-45,in=180] (0,0.05) to [out=0,in=45] (0.25,0.25);
 \draw [gray] (-0.5,0)--(0,-0.5)--(0.5,0)--(0,0.5)--cycle;
 \draw [gray,fill] (-0.25,0.25) circle(1pt);
 \draw [gray,fill] (-0.25,-0.25) circle(1pt);
 \draw [gray,fill] (0.25,0.25) circle(1pt);
 \draw [gray,fill] (0.25,-0.25) circle(1pt);
\end{tikzpicture} +
\begin{tikzpicture}[scale=0.85,baseline={([yshift=-3pt]current bounding box.center)}]
 \draw [thick,blue] (-0.25,0.25) -- (0.25,-0.25);
 \draw [gray] (-0.5,0)--(0,-0.5)--(0.5,0)--(0,0.5)--cycle;
 \draw [gray,fill] (-0.25,0.25) circle(1pt);
 \draw [gray,fill] (-0.25,-0.25) circle(1pt);
 \draw [gray,fill] (0.25,0.25) circle(1pt);
 \draw [gray,fill] (0.25,-0.25) circle(1pt);
\end{tikzpicture} +
\begin{tikzpicture}[scale=0.85,baseline={([yshift=-3pt]current bounding box.center)}]
 \draw [thick,blue] (-0.25,-0.25) -- (0.25,0.25);
 \draw [gray] (-0.5,0)--(0,-0.5)--(0.5,0)--(0,0.5)--cycle;
 \draw [gray,fill] (-0.25,0.25) circle(1pt);
 \draw [gray,fill] (-0.25,-0.25) circle(1pt);
 \draw [gray,fill] (0.25,0.25) circle(1pt);
 \draw [gray,fill] (0.25,-0.25) circle(1pt);
\end{tikzpicture} \right) + K^2 \left(
\begin{tikzpicture}[scale=0.85,baseline={([yshift=-3pt]current bounding box.center)}]
 \draw [thick,blue] (-0.25,-0.25) to [out=45,in=-90] (-0.05,0) to [out=90,in=-45] (-0.25,0.25);
 \draw [thick,blue] (0.25,-0.25) to [out=135,in=-90] (0.05,0) to [out=90,in=-135] (0.25,0.25);
 \draw [gray] (-0.5,0)--(0,-0.5)--(0.5,0)--(0,0.5)--cycle;
 \draw [gray,fill] (-0.25,0.25) circle(1pt);
 \draw [gray,fill] (-0.25,-0.25) circle(1pt);
 \draw [gray,fill] (0.25,0.25) circle(1pt);
 \draw [gray,fill] (0.25,-0.25) circle(1pt);
\end{tikzpicture} +
\begin{tikzpicture}[scale=0.85,baseline={([yshift=-3pt]current bounding box.center)}]
 \draw [thick,blue] (-0.25,-0.25) to [out=45,in=-180] (0,-0.05) to [out=0,in=135] (0.25,-0.25);
 \draw [thick,blue] (-0.25,0.25) to [out=-45,in=180] (0,0.05) to [out=0,in=45] (0.25,0.25);
 \draw [gray] (-0.5,0)--(0,-0.5)--(0.5,0)--(0,0.5)--cycle;
 \draw [gray,fill] (-0.25,0.25) circle(1pt);
 \draw [gray,fill] (-0.25,-0.25) circle(1pt);
 \draw [gray,fill] (0.25,0.25) circle(1pt);
 \draw [gray,fill] (0.25,-0.25) circle(1pt);
\end{tikzpicture} \right)
\end{equation}
is then an element of the dilute Jones Temperley-Lieb algebra $\dJTL$,   itself a periodicized version of the  dilute Temperley-Lieb algebra (isomorphic to the Motzkin algebra), denoted  $\dTL$ in \cite{JRS22}. In this diagram algebra, each fundamental vertex in (\ref{Fig6}) corresponds to a different generator. 
These generators can be depicted in a more conventional way by omitting the tile boundaries and deforming the monomer lines as follows:
\begin{equation}
\label{Fig6a}
\begin{tikzpicture}[scale=0.85,baseline={([yshift=-3pt]current bounding box.center)}]
 \draw [gray,fill] (-0.25,0.25) circle(1pt);
 \draw [gray,fill] (-0.25,-0.25) circle(1pt);
 \draw [gray,fill] (0.25,0.25) circle(1pt);
 \draw [gray,fill] (0.25,-0.25) circle(1pt);
\end{tikzpicture} \qquad\quad
\begin{tikzpicture}[scale=0.85,baseline={([yshift=-3pt]current bounding box.center)}]
 \draw [thick,blue] (-0.25,-0.25) to (-0.25,0.25);
 \draw [gray,fill] (-0.25,0.25) circle(1pt);
 \draw [gray,fill] (-0.25,-0.25) circle(1pt);
 \draw [gray,fill] (0.25,0.25) circle(1pt);
 \draw [gray,fill] (0.25,-0.25) circle(1pt);
\end{tikzpicture} \qquad\quad
\begin{tikzpicture}[scale=0.85,baseline={([yshift=-3pt]current bounding box.center)}]
 \draw [thick,blue] (0.25,-0.25) to (0.25,0.25);
 \draw [gray,fill] (-0.25,0.25) circle(1pt);
 \draw [gray,fill] (-0.25,-0.25) circle(1pt);
 \draw [gray,fill] (0.25,0.25) circle(1pt);
 \draw [gray,fill] (0.25,-0.25) circle(1pt);
\end{tikzpicture} \qquad\quad
\begin{tikzpicture}[scale=0.85,baseline={([yshift=-3pt]current bounding box.center)}]
 \draw [thick,blue] (-0.25,-0.25) to [out=90,in=90] (0.25,-0.25);
 \draw [gray,fill] (-0.25,0.25) circle(1pt);
 \draw [gray,fill] (-0.25,-0.25) circle(1pt);
 \draw [gray,fill] (0.25,0.25) circle(1pt);
 \draw [gray,fill] (0.25,-0.25) circle(1pt);
\end{tikzpicture} \qquad\quad
\begin{tikzpicture}[scale=0.85,baseline={([yshift=-3pt]current bounding box.center)}]
 \draw [thick,blue] (-0.25,0.25) to [out=-90,in=-90] (0.25,0.25);
 \draw [gray,fill] (-0.25,0.25) circle(1pt);
 \draw [gray,fill] (-0.25,-0.25) circle(1pt);
 \draw [gray,fill] (0.25,0.25) circle(1pt);
 \draw [gray,fill] (0.25,-0.25) circle(1pt);
\end{tikzpicture} \qquad\quad
\begin{tikzpicture}[scale=0.85,baseline={([yshift=-3pt]current bounding box.center)}]
 \draw [thick,blue] (-0.25,0.25) -- (0.25,-0.25);
 \draw [gray,fill] (-0.25,0.25) circle(1pt);
 \draw [gray,fill] (-0.25,-0.25) circle(1pt);
 \draw [gray,fill] (0.25,0.25) circle(1pt);
 \draw [gray,fill] (0.25,-0.25) circle(1pt);
\end{tikzpicture} \qquad\quad
\begin{tikzpicture}[scale=0.85,baseline={([yshift=-3pt]current bounding box.center)}]
 \draw [thick,blue] (-0.25,-0.25) -- (0.25,0.25);
 \draw [gray,fill] (-0.25,0.25) circle(1pt);
 \draw [gray,fill] (-0.25,-0.25) circle(1pt);
 \draw [gray,fill] (0.25,0.25) circle(1pt);
 \draw [gray,fill] (0.25,-0.25) circle(1pt);
\end{tikzpicture} \qquad\quad
\begin{tikzpicture}[scale=0.85,baseline={([yshift=-3pt]current bounding box.center)}]
 \draw [thick,blue] (-0.25,-0.25) to (-0.25,0.25);
 \draw [thick,blue] (0.25,-0.25) to (0.25,0.25);
 \draw [gray,fill] (-0.25,0.25) circle(1pt);
 \draw [gray,fill] (-0.25,-0.25) circle(1pt);
 \draw [gray,fill] (0.25,0.25) circle(1pt);
 \draw [gray,fill] (0.25,-0.25) circle(1pt);
\end{tikzpicture} \qquad\quad
\begin{tikzpicture}[scale=0.85,baseline={([yshift=-3pt]current bounding box.center)}]
 \draw [thick,blue] (-0.25,-0.25) to [out=90,in=90] (0.25,-0.25);
 \draw [thick,blue] (-0.25,0.25) to [out=-90,in=-90] (0.25,0.25);
 \draw [gray,fill] (-0.25,0.25) circle(1pt);
 \draw [gray,fill] (-0.25,-0.25) circle(1pt);
 \draw [gray,fill] (0.25,0.25) circle(1pt);
 \draw [gray,fill] (0.25,-0.25) circle(1pt);
\end{tikzpicture}
\end{equation}
(in fact, not all these generators are independent but this will not matter here \cite{JRS22}). On top of this we have the usual rules that loops get the fugacity $n$, and also that lines cannot stop---end points are not allowed. With periodic boundary conditions, we allow lines to connect around the system, and non-contractible loops are also given the fugacity $n$. It is important to understand that the algebra encodes all of the geometrical features, and is not dependent on the monomer fugacity $K$---the latter simply appears in the expression for the transfer matrix (or Hamiltonian) in the form of coefficients of words in the algebra, as in (\ref{Tmat}). 

The defect line $D$ is then introduced as shown in figure \ref{Fig7}: we add a line of defect tiles carrying a red line (or  non-contractible loop on the cylinder). There are two types of defect tiles that alternate throughout a horizontal
row of the system:
\begin{equation}
\label{fig:defect-tile}
\begin{tikzpicture}[scale=1,baseline={([yshift=-3pt]current bounding box.center)}]
 \draw [fill,blue!10] (0,0) -- (0.5,0.5) -- (0.5,1.0) --  (0,0.5) -- cycle; 
 \draw [very thick,black] (0,0) -- (0.5,0.5) -- (0.5,1.0) -- (0,0.5) -- cycle;
 \draw [thick,red] (0,0.25) -- (0.5,0.75);
 \draw [gray,fill] (0.25,0.25) circle(1pt);
 \draw [gray,fill] (0.25,0.75) circle(1pt);
\end{tikzpicture} =
\begin{tikzpicture}[scale=1,baseline={([yshift=-3pt]current bounding box.center)}]
 \draw [thick,red] (0,0.25) -- (0.5,0.75);
 \draw [thick,blue] (0.25,0.25) -- (0.25,0.45);
 \draw [thick,blue] (0.25,0.55) -- (0.25,0.75);
 \draw [very thick,black] (0,0) -- (0.5,0.5) -- (0.5,1.0) -- (0,0.5) -- cycle;
 \draw [gray,fill] (0.25,0.25) circle(1pt);
 \draw [gray,fill] (0.25,0.75) circle(1pt);
\end{tikzpicture} +
\begin{tikzpicture}[scale=1,baseline={([yshift=-3pt]current bounding box.center)}]
 \draw [thick,red] (0,0.25) -- (0.5,0.75);
 \draw [very thick,black] (0,0) -- (0.5,0.5) -- (0.5,1.0) -- (0,0.5) -- cycle;
 \draw [gray,fill] (0.25,0.25) circle(1pt);
 \draw [gray,fill] (0.25,0.75) circle(1pt);
\end{tikzpicture} \qquad \mbox{and} \qquad
\begin{tikzpicture}[scale=1,baseline={([yshift=-3pt]current bounding box.center)}]
 \draw [fill,blue!10] (0,0) -- (0.5,-0.5) -- (0.5,-1.0) --  (0,-0.5) -- cycle; 
 \draw [thick,red] (0,-0.25) -- (0.5,-0.75);
 \draw [very thick,black] (0,0) -- (0.5,-0.5) -- (0.5,-1.0) -- (0,-0.5) -- cycle;
 \draw [gray,fill] (0.25,-0.25) circle(1pt);
 \draw [gray,fill] (0.25,-0.75) circle(1pt);
\end{tikzpicture} =
\begin{tikzpicture}[scale=1,baseline={([yshift=-3pt]current bounding box.center)}]
 \draw [thick,red] (0,-0.25) -- (0.5,-0.75);
 \draw [thick,blue] (0.25,-0.25) -- (0.25,-0.45);
 \draw [thick,blue] (0.25,-0.55) -- (0.25,-0.75);
 \draw [very thick,black] (0,0) -- (0.5,-0.5) -- (0.5,-1.0) -- (0,-0.5) -- cycle;
 \draw [gray,fill] (0.25,-0.25) circle(1pt);
 \draw [gray,fill] (0.25,-0.75) circle(1pt);
\end{tikzpicture} +
\begin{tikzpicture}[scale=1,baseline={([yshift=-3pt]current bounding box.center)}]
 \draw [thick,red] (0,-0.25) -- (0.5,-0.75);
 \draw [very thick,black] (0,0) -- (0.5,-0.5) -- (0.5,-1.0) -- (0,-0.5) -- cycle;
 \draw [gray,fill] (0.25,-0.25) circle(1pt);
 \draw [gray,fill] (0.25,-0.75) circle(1pt);
\end{tikzpicture}
\end{equation}
but actually they differ only in the way we draw them, namely by the alternating inclination of their
bottom and top boundaries with respect to horizontal. More importantly, their algebraic expansions
are identical, with
the two terms on the right-hand side fitting, respectively, ordinary tiles with and without monomers
that can be placed below and above the defect tile so as to touch its grey points.
Nothing happens if the defect tile is empty of
monomers (second term), otherwise monomers can go below $D$ (first term)---the meaning of ``going below'' being defined by the decomposition (\ref{Fig1}). Instead of below, we could decide that monomers go above the defect instead,
giving rise to a different defect line $\overline{D}$. 

\subsection{The dense case}

The fully dense (also called ``completely packed'' in some works) case%
\footnote{We call here ``fully dense'' the case  where the monomer fugacity is very large, so that all edges of the lattice $S$ are occupied. The model then coincides with the model of ``surrounding lattice decompositions'' that appears, e.g., in the reformulation of the Potts model \cite{BKW76}. It is widely believed that our lattice model for all $K>K_{\rm c}$ is in the same universality class as this fully dense limit. We use the name fully dense in contrast with the often used ``fully packed'' to stress the fact that loop models with maximum occupancy of the edges and/or vertices enjoy less universality than their dense but not maximally so counterparts. On some lattices, the thermodynamic limit and the $K\to\infty$ limits do not commute \cite{KGN96,JK98}. We shall not be concerned with this subtlety here.}
occurs when the monomer fugacity is very large ($K \to \infty$), so all edges of the lattice $S$ are occupied.%
\footnote{Note that in that limit, it does not matter whether all fugacities are the same or not: provided they are taken to be large, the ``monomer'' weight of all configurations is the same---the monomer fugacity couples to a redundant operator.}
In this case, the relevant algebra is the unoriented Jones Temperley-Lieb algebra $\uJTL_{\!\!N}(\n)$. 
 Correspondingly, the ordinary tiles are restricted to the last two diagrams of (\ref{Fig6}), and the defect tile corresponds to only the first term on the right-hand sides of (\ref{fig:defect-tile}). 

The fully dense case has in fact already been studied in \cite{BGJST}. 
It was shown in this paper that the two defects $\DD$ and $\overline{\DD}$ corresponding to passing a loop over or under the other lines commute with the whole algebra. It is easy, in fact, to write expressions for these defects  by systematically splitting the crossings. One finds
\begin{eqnarray}
\DD&=&(-q)^{-N/2}\tau(1-qe_{N-1})\cdots(1-qe_2)(1-qe_1)+(-q)^{N/2}(1-q^{-1}e_1)\cdots(1-q^{-1}e_{N-1})\tau^{-1} \,,\nonumber\\
\overline{\DD}&=&(-q)^{N/2}\tau(1-q^{-1}e_{N-1})\cdots(1-q^{-1}e_2)(1-q^{-1}e_1)+(-q)^{-N/2}(1-qe_1)\cdots(1-qe_{N-1})\tau^{-1} \,, \nonumber
\end{eqnarray}
where $N$ is the width of the system,  and $\tau$ translates every site one step to the right. The main result of \cite{BGJST} can then be formulated as\footnote{In fact, reference \cite{BGJST} also addresses the question of whether these are the only operators satisfying this property---in more precise terms, the center of the algebra. We will not discuss this aspect here.}
\begin{equation}
\left[\DD,\uJTL_{\!\!N}(\n)\right]=\left[\overline{\DD},\uJTL_{\!\!N}(\n)\right]=0 \,.
\end{equation}
These commutation relations mean that the defect operators commute with all Hamiltonians built out of elements of the algebra. They also of course commute with local  (as well as row-to-row) transfer matrices, implying in particular invariance of the partition function of the model under arbitrary deformations of  the associated TDL $D$. We insist on the fact that this result is true on the lattice, before any kind of continuum limit is taken. 

While it is convenient to think of this problem geometrically, it is important to realize that all we said has a strict equivalent, when $n$ is an integer, for the original  $O(n)$ spin model. The Hamiltonian or transfer matrix acts on the tensor product of fundamental representations ${\cal H}=[1]^{\otimes N}$, with the well-known actions (here and later the Roman labels $a$, $b$ etc.\ take values $1,2,\ldots,n$ and all sums run over this set of values):
\begin{subequations}
\begin{eqnarray}
e_i |a_1,\ldots,a_N\rangle&=&\delta_{a_ia_{i+1}}\sum_{b} |a_1,\ldots,a_{i-1},b,b,a_{i+2},\ldots a_N\rangle \,, \\
\tau |a_1,\ldots,a_N\rangle&=&|a_N,a_1,\ldots,a_{N-1}\rangle \,.
\end{eqnarray}
\end{subequations}
These formulas give rise to an explicit form for the operators $\DD,\overline{\DD}$. On  two ($N=2$) and three ($N=3$) sites, for instance, we have
\begin{subequations}
\begin{eqnarray}
\DD|ab\rangle \!\!\! &=& \!\! -(q+q^{-1})|ba\rangle+2\delta_{ab}\sum_c|cc\rangle \,, \label{2example} \\
\DD|abc\rangle \!\!\! &=& \!\! (-q)^{-3/2}|cab\rangle+(-q)^{3/2}|bca\rangle+\delta_{ab}\big((-q)^{1/2}\sum_d |dcd\rangle+(-q)^{-1/2}\sum_d |cdd\rangle\big) \label{3example} \\
&+& \!\! \delta_{bc}\big((-q)^{1/2}\sum_d|dda\rangle+(-q)^{-1/2}\sum_d |dad\rangle\big)
+\delta_{ac}\big((-q)^{1/2}\sum_d |bdd\rangle+(-q)^{-1/2}\sum_d|ddb\rangle\big) \,. \nonumber
\end{eqnarray}
\end{subequations}
Since $\DD$ is expressed in terms of the Temperley-Lieb generator and the shift operator, it commutes with $O(n)$
 in the sense of quantum Shuhr-Weyl duality:
\begin{equation}
\left[\DD,O(n)\right]=0 \,.
\end{equation}
Indeed, the space of states of the fully dense $O(n)$ model can be written as a bimodule over $\uJTL\otimes O(n)$:
\begin{align}
 \mathcal{H}_{O(n)} = W_{\langle 1,1\rangle}\otimes [\,] \oplus \bigoplus_{r\in\frac12\mathbb{N}^*}\bigoplus_{\substack{s\in\frac{1}{r}\mathbb{Z}\\ -1<s\leq 1}} W_{(r,s)}\otimes  \Lambda_{(r,s)}\ .
 \label{soni}
\end{align}
Here the $\Lambda_{(r,s)}$ are sums of $O(n)$ representations, a few of which are \cite{JRS22}
\begingroup
\allowdisplaybreaks
\begin{subequations}
\label{lxxx-all}
\begin{align}
\Lambda_{(\frac12,0)} &= [1]\ , \label{lhz}
 \\
 \Lambda_{(1,0)} &= [2]\ ,
 \\
 \Lambda_{(1,1)} &= [1^2]\ , \label{loo}
 \\
\Lambda_{(\frac32,0)}&= [3]+[1^3]\ ,
\label{l320}
\\
\Lambda_{(\frac32,\frac23)} &= [21]\ ,
\label{l3223}
\\
\Lambda_{(2,0)}&= [4]+[2^2]+[21^2]+[2]+[\,]\ ,
\label{l20}
\\
\Lambda_{(2,\frac12)}& = [31]+[21^2]+[1^2]\ ,
\\
\Lambda_{(2,1)} &= [31]+[2^2]+[1^4]+[2]\ .
\label{l21}
\end{align}
\end{subequations}
\endgroup
These result still hold for $n\in\mathbb{C}$, with a proper definition of the $O(n)$ symmetry for $n$ non integer, as discussed in \cite{JRS22,BR19}. The objects $W_{\langle 1,1\rangle}$ and $W_{(r,s)}$ appearing in (\ref{soni}) are irreducible modules of $\uJTL$. Here, the label $r$ is one-half  the number of non-contractible lines, while the label $s$ is a pseudomomentum that arises from representations of the cyclic group acting on these lines. Geometrically, it can be understood from  phase terms $z$  that are gained by each individual line when it winds around the cylinder. It is customary to set $z=e^{i\pi s}$, with the condition $z^{2r}=e^{2i\pi rs}=1$ that a simultaneous winding of all the lines produces no phase. The eigenstate with the smallest eigenvalues in $W_{(r,s)}$ gives  the  leading term in the two-point function of the corresponding generalized  fuseau operators. The module $W_{\langle 1,1\rangle}$ (sometimes denoted by $\overline{W}_{0,q^{\pm 2}}$ in other work) is the module with no throughlines,  where  non-contractible loops can occur. Since they are given the same weight  $n$ as contractible ones, the way two points are connected in this representation  of the algebra does not matter. 
In this setup, $\Lambda_{(2,0)}$ is then the $O(n)$ multiplicity space of the four-leg operator without winding phases which was mentioned in the Introduction.

Every term in the sum (\ref{soni}) is an eigenspace (for $q$ not a root of unity\footnote{Subtleties arise when $q$ is a root of unity, where some representations might not be fully reducible, and operators not diagonalizable.}) of the operators $\DD,\overline{\DD}$. The corresponding eigenvalues are
\begin{subequations}
\label{eigen}
\begin{eqnarray}
\DD_{\langle 11\rangle}&=&q+q^{-1} \,, \\
\DD_{(r,s)}&=&e^{i\pi s}(-q)^r+e^{-i\pi s}(-q)^{-r} \,,
\end{eqnarray}
\end{subequations}
and the same with $q\to q^{-1}$ for $\overline{\DD}$. For $q$ generic, it is easy to see that the eigenvalues of the $\DD$ operators determine the $(r,s)$ uniquely. Note that we can think of the different $O(n)$ representations involved in a given $\Lambda_{(r,s)}$ as giving rise to the same eigenvalue of $\DD,\overline{\DD}$. Going back to $n$ integer and taking $a\neq b\neq c$ we have, for instance,
\begin{subequations}
\begin{eqnarray}
|abc\rangle_{[3]}&\equiv&|abc\rangle+|acb\rangle+|bac\rangle+|bca\rangle+|cba\rangle+|cab\rangle \,, \\
|abc\rangle_{[1]^3}&\equiv&|abc\rangle-|acb\rangle-|bac\rangle+|bca\rangle-|cba\rangle+|cab\rangle \,,
\end{eqnarray}
\end{subequations}
and 
\begin{subequations}
\begin{eqnarray}
\DD|abc\rangle_{[3]}&=&((-q)^{-3/2}+(-q)^{3/2}) |abc\rangle_{[3]} \,, \\
\DD|abc\rangle_{[1]^3}&=&((-q)^{-3/2}+(-q)^{3/2}) |abc\rangle_{[111]} \,,
\end{eqnarray}
\end{subequations}
with the same eigenvalue---a result compatible with the identity $\Lambda_{({3\over 2},0)}=[3]+[1]^3$ from (\ref{l320}).

We note that it is possible to build ``higher-spin'' defects by fusion. In the integrable  construction, the defect line can be  considered as carrying a spin-$1/2$ representation of $U_qsl(2)$: it is perfectly possible to equip it with a spin-$j$ representation instead (for any $2j \in \mathbb{N}^*$). This can be done in practice by introducing $2j$ defect lines, and projecting them onto spin $j$ using a Jones-Wenzl projector. The construction has been discussed in detail in \cite{BGJST} and leads to the straightforward identity
\begin{equation}
\DD^{(j)} \DD=\DD^{(j+1/2)}+\DD^{(j-1/2)} \,,\label{higherdef}
\end{equation}
with $\DD=\DD^{(1/2)}$. Hence we have in particular
\begin{subequations}
\begin{eqnarray}
\DD^{(1)}&=&\DD^2-1 \,, \\
\DD^{(3/2)}&=&\DD^3-2\DD \,, \\
\DD^{(2)}&=&\DD^4-3\DD^2+1 \,.
\end{eqnarray}
\end{subequations}
In practice  we shall not need these higher defects in our analysis: although their existence is important for some other purposes, we shall  not consider them further here. 
\subsection{The dilute case}

The construction obviously generalizes to the case where not every edge of the lattice $S$  is covered by a monomer: in this case, there is no interaction with  the defect line, while, if there is a monomer, the interaction is as before. Here is an explicit example:
\begin{eqnarray}
\label{Fig9}
 {\sf D} \
 \begin{tikzpicture}[scale=0.7,baseline={([yshift=-3pt]current bounding box.center)}]
 \draw [dashed,black] (0,0) -- (0,1);
 \draw [thick,blue] (1,0) -- (1,1);
 \draw [dashed,black] (2,0) -- (2,1);
 \draw [thick,blue] (3,0) -- (3,1);
\end{tikzpicture} &=&
 \begin{tikzpicture}[scale=0.7,baseline={([yshift=-3pt]current bounding box.center)}]
 \draw [dashed,black] (0,0) -- (0,1);
 \draw [thick,blue] (1,0) -- (1,0.45);
 \draw [thick,blue] (1,0.55) -- (1,1);
 \draw [dashed,black] (2,0) -- (2,1);
 \draw [thick,blue] (3,0) -- (3,0.45);
 \draw [thick,blue] (3,0.55) -- (3,1);
 \draw [thick,red] (-0.5,0.5) -- (3.5,0.5);
\end{tikzpicture} \nonumber \\
&=& (-q)
 \begin{tikzpicture}[scale=0.7,baseline={([yshift=-3pt]current bounding box.center)}]
 \draw [dashed,black] (0,0) -- (0,1);
 \draw [thick,blue] (1,0) -- (1,0.3) arc(0:45:2mm);
 \draw [thick,blue] (1,1) -- (1,0.69) arc(180:225:2mm);
 \draw [dashed,black] (2,0) -- (2,1);
 \draw [thick,blue] (3,0) -- (3,0.3) arc(0:45:2mm);
 \draw [thick,blue] (3,1) -- (3,0.69) arc(180:225:2mm);
 \draw [thick,red] (-0.5,0.5) -- (0.8,0.5) arc(90:45:2mm);
 \draw [thick,red] (1.05,0.55) arc(225:270:2mm) -- (2.8,0.5) arc(90:45:2mm);
 \draw [thick,red] (3.05,0.55) arc(225:270:2mm) -- (3.5,0.5);
\end{tikzpicture} + (-q)^{-1}
 \begin{tikzpicture}[scale=0.7,baseline={([yshift=-3pt]current bounding box.center)}]
 \draw [dashed,black] (0,0) -- (0,1);
 \draw [thick,blue] (1,0) -- (1,0.3) arc(180:135:2mm);
 \draw [thick,blue] (1,1) -- (1,0.69) arc(0:-45:2mm);
 \draw [dashed,black] (2,0) -- (2,1);
 \draw [thick,blue] (3,0) -- (3,0.3) arc(180:135:2mm);
 \draw [thick,blue] (3,1) -- (3,0.69) arc(0:-45:2mm);
 \draw [thick,red] (-0.5,0.5) -- (0.8,0.5) arc(270:315:2mm);
 \draw [thick,red] (1.06,0.445) arc(135:90:2mm) -- (2.83,0.5) arc(280:315:2mm);
 \draw [thick,red] (3.06,0.445) arc(135:90:2mm) -- (3.5,0.5);
\end{tikzpicture} \nonumber \\
& & +
 \begin{tikzpicture}[scale=0.7,baseline={([yshift=-3pt]current bounding box.center)}]
 \draw [dashed,black] (0,0) -- (0,1);
 \draw [thick,blue] (1,0) -- (1,0.3) arc(0:45:2mm);
 \draw [thick,blue] (1,1) -- (1,0.69) arc(180:225:2mm);
 \draw [dashed,black] (2,0) -- (2,1);
 \draw [thick,blue] (3,0) -- (3,0.3) arc(180:135:2mm);
 \draw [thick,blue] (3,1) -- (3,0.69) arc(0:-45:2mm);
 \draw [thick,red] (-0.5,0.5) -- (0.8,0.5) arc(90:45:2mm);
 \draw [thick,red] (1.05,0.55) arc(225:270:2mm) -- (2.83,0.5) arc(280:315:2mm);
 \draw [thick,red] (3.06,0.445) arc(135:90:2mm) -- (3.5,0.5);
\end{tikzpicture} +
 \begin{tikzpicture}[scale=0.7,baseline={([yshift=-3pt]current bounding box.center)}]
 \draw [dashed,black] (0,0) -- (0,1);
 \draw [thick,blue] (1,0) -- (1,0.3) arc(180:135:2mm);
 \draw [thick,blue] (1,1) -- (1,0.69) arc(0:-45:2mm);
 \draw [dashed,black] (2,0) -- (2,1);
 \draw [thick,blue] (3,0) -- (3,0.3) arc(0:45:2mm);
 \draw [thick,blue] (3,1) -- (3,0.69) arc(180:225:2mm);
 \draw [thick,red] (-0.5,0.5) -- (0.8,0.5) arc(270:315:2mm);
 \draw [thick,red] (1.06,0.445) arc(135:90:2mm) -- (2.8,0.5) arc(90:45:2mm);
 \draw [thick,red] (3.05,0.55) arc(225:270:2mm) -- (3.5,0.5);
\end{tikzpicture}
\end{eqnarray}
In terms of $O(n)$ variables, the defect operator acts on states now in $\left([\,]+[1]\right)^{\otimes N}$ by being simply transparent to the variables in the trivial representation $[\,]$ (geometrically an empty edge), while acting on those in the fundamental $[1]$ (an edge carrying a blue line) as before. So, for instance, corresponding to (\ref{Fig9}) we have
\begin{equation}
\DD|0a0b\rangle=-(q+q^{-1})|0b0a\rangle+2\delta_{ab}\sum_c |0c0c\rangle \,,
\end{equation}
to be compared with (\ref{2example}).  The Roman labels here take the same values $1,2,\ldots,n$ as before, whereas $0$ refers to the empty state in the trivial representation $[\,]$. The Hilbert space  still decomposes as in (\ref{soni}), but now with representations of the dilute algebra  $\dJTL$:
\begin{align}
 \mathcal{H}_{O(n)} = \widetilde{W}_{\langle 1,1\rangle}\otimes [\,] \oplus \bigoplus_{r\in\frac12\mathbb{N}^*}\bigoplus_{\substack{s\in\frac{1}{r}\mathbb{Z}\\ -1<s\leq 1}} \widetilde{W}_{(r,s)}\otimes  \Lambda_{(r,s)}\ .
 \label{sonii}
\end{align}
In contrast with the dense case, the monomer fugacity affects non-trivially the properties (partition function and correlation functions) of the model. But it does {\sl not} affect the formal  structure of the Hilbert space decomposition (\ref{sonii}). The defect operators are still diagonal on each term  in the sum, with eigenvalues still given by (\ref{eigen}).

\section{The non-invertible  symmetry at the dilute fixed point and in the RG flow}

At the dilute critical point, the decomposition (\ref{soni}) carries over to the continuum limit, and hence holds for the fixed-point CFT as well. Each  $\dJTL$ module becomes a sum over representations of $\VirN$:
\begin{subequations}
\begin{eqnarray}
\widetilde{W}_{\langle 1,1\rangle} &\mapsto& \bigoplus_{s\in 2\mathbb{N}+1} \mathcal{R}_{\langle 1,s\rangle}  \,, \\
\widetilde{W}_{(r,s)} &\mapsto& \bigoplus_{s'\in 2\mathbb{Z}+s} \mathcal{W}_{(r,s)} \,,
\end{eqnarray}
\end{subequations}
where the notations for the representations and their content of primaries are as follows \cite{JRS22}:
\begin{align}
\renewcommand{\arraystretch}{1.5}
\renewcommand{\arraycolsep}{8pt}
 \begin{array}{|l|l|l|l|}
 \hline
 \text{Notation} & \text{Indices} & \text{Name} & \text{Primary states}
 \\
 \hline\hline
  \mathcal{R}_{\langle 1,s\rangle} & s\in \mathbb{N}^* &  \text{Degenerate} & (\Delta_{(1,s)},\Delta_{(1,s)})
  \\
  \hline
  \mathcal{W}_{(r,s)} & r\notin\mathbb{Z}^*\text{ or } s\notin\mathbb{Z}^* & \text{Verma module} & (\Delta_{(r,s)},\Delta_{(r,-s)})
  \\
  \hline 
  \mathcal{W}_{(r,s)} & r,s\in\mathbb{N}^* & \text{Logarithmic} & 
  \renewcommand{\arraystretch}{1}
  \begin{array}{@{}l@{}}  (\Delta_{(r,s)},\Delta_{(r,-s)}) \\ (\Delta_{(r,-s)},\Delta_{(r,s)}) \\ (\Delta_{(r,-s)},\Delta_{(r,-s)}) \end{array} 
  \renewcommand{\arraystretch}{1.5}
  \\
  \hline
  \mathcal{W}_{(r,s)} & r,-s\in\mathbb{N}^* & 0 & 
  \\
  \hline
 \end{array}
 \label{vwww}
\end{align}

Rather than dwell on the (irrelevant for us) details of these representations, let us give the corresponding characters (the superscript $N$ stands for non-diagonal):
\begin{subequations}
\label{chis}
\begin{eqnarray}
 \chi_{\langle 1,s\rangle}(\tau) &=& \left| \frac{e^{2\pi i\tau P^2_{(1,s)}} - e^{2\pi i \tau P^2_{(1,-s)}}}{\eta(\tau)}\right|^2 \,, \\
 \chi^N_{(r,s)}(\tau) &=& \frac{e^{2\pi i\tau P^2_{(r,s)}} e^{2\pi i\bar\tau P^2_{(r,-s)}}}{|\eta(\tau)|^2} \,,
\end{eqnarray}
\end{subequations}
where $\eta(\tau)$ is the Dedekind eta function.  We have then
\begin{subequations}
\begin{align}
 \operatorname{Tr}_{\mathcal{R}_{\langle 1,s\rangle}}\left( e^{2\pi i \tau(L_0-\frac{c}{24})} e^{2\pi i \bar\tau(\bar L_0-\frac{c}{24})}\right) &= \chi_{\langle 1,s\rangle}(\tau)\ ,
 \\
 \renewcommand{\arraystretch}{2.5}
  \operatorname{Tr}_{\mathcal{W}_{(r,s)}}\left( e^{2\pi i \tau(L_0-\frac{c}{24})} e^{2\pi i \bar\tau(\bar L_0-\frac{c}{24})}\right) &= \left\{\begin{array}{ll} \chi^N_{(r,s)}(\tau) & \text{if} \quad r\notin \mathbb{Z}^*\text{ or } s\notin\mathbb{Z}^*\ ,                                                                                                                                                             \\ 
         \chi^N_{(r,s)}(\tau) + \chi^N_{(r,-s)}(\tau) & \text{if} \quad r,s\in\mathbb{N}^*\ ,
         \\ 0 & \text{if} \quad r,-s\in\mathbb{N}^*\ .
         \end{array}\right.
\end{align}
\end{subequations}
Recall also the central charge 
\begin{equation}
c=13-6\beta^2-{6\over \beta^2}
\end{equation}
together with the parametrization
\begin{equation}
\label{nparam}
n=-2\cos (\pi\beta^2) \,, \quad \beta^2\in [1,2]
\end{equation}
($\beta^2$ is often denoted by the Coulomb gas coupling constant $g$ in the $O(n)$-model literature). Finally we have set
\begin{equation}
\label{Pparam}
P_{(r,s)}={1\over 2}(\beta r-\beta^{-1}s) \,,
\end{equation}
so the conformal weights are given by 
\begin{equation}
\label{confweights}
\Delta_{(r,s)}=P_{(r,s)}^2-P_{(1,1)}^2 \,.
\end{equation}

Since the topological nature of the defect line is exact on the lattice for all finite sizes, it holds as well in the continuum limit. In this limit, the  $\dJTL$ algebra becomes the interchiral algebra $ \widetilde{\mathfrak{C}}_{\beta^2}$, an algebra containing $\VirN$ (denoted  $\mathfrak{C}_c$ in \cite{JRS22}) but extended by the degenerate field $V_{\langle 1,3\rangle}$:
\begin{align}
 \mathfrak{C}_c = \text{Span}\left(T, \bar T\right) \quad \subset  \quad \widetilde{\mathfrak{C}}_{\beta^2} = \text{Span}\left(T,\bar T, V_{\langle 1, 3\rangle}\right)  \,,
\end{align}
\begin{align}
 \lim_{\text{critical}} \dJTL(n) = \widetilde{\mathfrak{C}}_{\beta^2} \,. 
\end{align}
We see that the topological defects in fact commute not only with  $\mathfrak{C}_c$ but with this extended algebra as well:
\begin{equation}
\left[\DD, \widetilde{\mathfrak{C}}_{\beta^2}\right]=0 \,,
\end{equation}
where we have denoted by the same symbol $\DD$ the defect on the lattice and in the continuum limit of the dilute critical point.  

Now let us imagine perturbing this critical point by some interaction on the lattice, leading to a new  Hamiltonian $H_{\rm pert}$. As long as this interaction can be expressed locally in terms of the elements in  $\dJTL$ (which is the case for the thermal perturbation), it will give rise, in the  continuum limit, to  an expansion in terms of conformal fields that occur in the continuum limit of the  $\dJTL$ {\sl identity module}---which is what we have denoted $\mathcal{R}_{\langle 1,s\rangle}$. This is because every word  $\mathsf{w}$ in generators of the (dilute) algebra $\dJTL$ acting on the ground state of the theory will produce a state in $\widetilde{W}_{\langle 1,1\rangle}$: $\mathsf{w}|\mathds{1}\rangle=\sum_\alpha |\mathsf{w}_\alpha\rangle$---since this is a representation space of the algebra, and since it contains the ground state. By the state-operator correspondence, each state $|\mathsf{w}_\alpha\rangle$ will then  correspond to an  operator within  $\mathcal{R}_{\langle 1,s\rangle}$. Hence we have
\begin{equation}
\mathsf{w}\approx \sum_\alpha c_\alpha \Phi_{\alpha} \,,
\end{equation}
where the sum is over fields of the CFT in $\mathcal{R}_{\langle 1,s\rangle}$---i.e.\ descendants  under $\VirN$ of the primary fields $\Phi_{\langle 1,s\rangle}$---and the $c_\alpha$ are non-universal constants depending on the choice of lattice and the particular choice of microscopic model (as well as on the lattice spacing). It follows that, for any such perturbation expressed locally in terms of $\dJTL$, the resulting Hamiltonian still commutes with the topological defect, $[H_{\rm pert},\DD]=0$. 

Now the point is that the four-leg operator does not belong to $\mathcal{R}_{\langle 1,s\rangle}$. This is of course obvious from the fact that its eigenvalue \eqref{eigen} under the defect operator is different from the one of the ground state (for $q$ generic). In fact, classifying fields using their  associated  $\dJTL$ module is equivalent to classifying them in terms of their $\DD$ eigenvalues. Hence, once we know that the perturbation is expressed in terms of $\dJTL$, this guarantees that  the four-leg operator cannot appear  in the continuum limit of $H_{\rm pert}$. Moreover, fields in the identity module are stable under operator product expansion---this is known explicitly from the fact that all conformal fields in the $\mathcal{R}_{\langle 1,s\rangle}$ are exactly degenerate. It also follows from the topological argument: insertion of several fields in the identity sector can always be represented on the lattice as a word $\mathsf{w}|\mathds{1}\rangle$, and thus again decomposes onto
fields in the identity sector. Hence the four-leg operator never appears when doing the RG for $H_{\rm pert}$, and  cannot drive the  flow away from the known, topological symmetry-preserving, IR dense fixed point. 

We note that a similar argument was made in \cite{Trebstetal} to argue topological protection of ``topological phases'' in anyonic chains (see also \cite{BuicanGromov}). In fact, the  anyonic operator $Y$ of \cite{Trebstetal} coincides formally with $\DD$ once formulated in terms of the  Temperley-Lieb algebra. In this correspondence, however, the critical point considered in \cite{Trebstetal}  coincides with  the dense critical point  of the $O(n)$ model---the situation is  thus physically  different from the one we are interested in here.

The four-leg operator is prevented from appearing not by $O(n)$ symmetry (since it can occur within the singlet representation, as argued in the Introduction) but by the ``non-invertible symmetry'' which is a more elaborate way to say that lattice loops are prevented from crossing. We notice that it is generally believed that allowing crossings {\sl at the critical point} would not change the universality class---in agreement with the fact that the four-leg operator at this fixed point is irrelevant (this is  discussed more fully in Appendix~\ref{app:A}).

In the case with crossings, however, the non-invertible symmetry would not be present {\sl on the lattice}, since the four-leg operator does {\sl not} commute with the $\dJTL$ algebra, and, in its presence, the defect line cannot be moved around any longer. Here is an example (where we call $\Pi_{12}$ the operator crossing lines 1 and 2): we have
%
%
\begin{eqnarray}
\DD \Pi_{12}|abc\rangle=(-q)^{-3/2}|cba\rangle+(-q)^{3/2}|acb\rangle+\delta_{ab}\left((-q)^{1/2}\sum_d |dcd\rangle+(-q)^{-1/2}\sum_d |cdd\rangle\right)\nonumber\\
+\delta_{ac}\left((-q)^{1/2}\sum_d|ddb\rangle+(-q)^{-1/2}\sum_d |dbd\rangle\right)
+\delta_{bc}\left((-q)^{1/2}\sum_d |add\rangle+(-q)^{-1/2}\sum_d|dda\rangle\right) \nonumber
\end{eqnarray}
and%
\footnote{The reader may notice that these two expressions only differ by $q \to q^{-1}$, implying that $\DD \Pi_{12} = \Pi_{12} \overline{\DD}$ in this example. This identity is however a coincidence for this small size, and does not hold for larger $N$, as can be verified by going through the same computation for $N=4$.}
\begin{eqnarray}
\Pi_{12}\DD|abc\rangle=(-q)^{-3/2}|acb\rangle+(-q)^{3/2}|cba\rangle+\delta_{ab}\left((-q)^{1/2}\sum_d |cdd\rangle+(-q)^{-1/2}\sum_d |dcd\rangle\right)\nonumber\\
+\delta_{bc}\left((-q)^{1/2}\sum_d|dda\rangle+(-q)^{-1/2}\sum_d |add\rangle\right)
+\delta_{ac}\left((-q)^{1/2}\sum_d |dbd\rangle+(-q)^{-1/2}\sum_d|ddb\rangle\right) \,. \nonumber
\end{eqnarray}

With crossings allowed (giving rise to what is sometimes called  self-avoiding trails  \cite{OP} when $n=0$), the dilute critical point remains the same. In this case, therefore, the degeneracies in the partition function appear only in the continuum limit, and so does the topological symmetry. Increasing  the fugacity of monomers  is now a perturbation that will generically involve a component on the four-leg operator, since the lattice model is not topologically protected: the flow will then look as in  figure \ref{Fig0}. 

\section{The defect Hilbert space}

\begin{figure}
\centering
\begin{tikzpicture}[scale=0.69,baseline={([yshift=-3pt]current bounding box.center)}]
 \draw [thick,blue] (1,1) arc(-120:229:0.75 and 0.75);
 \draw [thick,red] (-1.3,-1.3) -- (1.375,1.5);
 \draw [thick,black] (1.275,1.5) -- (1.475,1.5);
 \draw [thick,black] (1.375,1.4) -- (1.375,1.6);
 \draw [thick,blue] (0.3,0.50) arc(40:190:0.5 and 1.2);
 \draw [thick,blue] (-0.55,-0.6) arc(210:375:0.5 and 1.2);
 \draw [thick,black] (-1.4,-1.3) -- (-1.2,-1.3);
 \draw [thick,black] (-1.3,-1.4) -- (-1.3,-1.2);
\end{tikzpicture}
\caption{The insertion of a pair of disorder fields  affects  loops encircling the extremities of the corresponding defect line.}
\label{Fig12}
\end{figure}
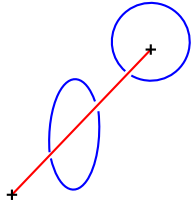

Consider now the situation where we  have an open version of our defect line, which we take to start at the origin and end at the point $(z,\bar{z})$.  By definition, the $O(n)$ model loops will then pass ``under'' this defect line, as shown in figure~\ref{Fig12}. This means that only loops encircling one (and only one) of the two end-points of the defect line will be affected by its presence. By decomposing the corresponding crossings as usual, using (\ref{Fig1}), we see that we obtain essentially a one leg-operator, but care must be taken. On the one hand, any given loop encircling one of the extremities has, before decomposition of the crossings, a fugacity $n$ that disappears after decomposition; on the other hand, each term in the decomposition comes with a factor $(-q)^{\pm 1/2}$. 

To handle these two facts properly, it is convenient to map the system onto the cylinder. We are then in a situation where the TDL,  instead of acting as the operator  $\DD$, runs parallel to the imaginary time direction and gives rise to a modified ``defect'' Hilbert space \cite{CSLWY}.  In the absence of through lines and of defects, the weight $n$ for non-contractible loops can be  obtained by orienting these loops,  summing over the two possible orientations, and giving oriented lines that go around the system a factor $z$ or $z^{-1}$. To get the correct loop fugacity $n = q + q^{-1}$, we should set $z=q$. 
 
In the presence of the  TDL  we decompose the crossings and see now that the problem splits into two sectors, one with a new parameter $z_1=(-q)^{1/2}z$ and the other with $z_2=(-q)^{-1/2}z$. Moreover, there is now {\sl one} throughline traversing the system. 
 
Choosing then $z=q$, we see that we get $z_1=-(-q)^{3/2}$ and $z_2=-(-q)^{1/2}$. For a general value $z=e^{i\pi e_\phi}$, the associated conformal weights are well known from the analysis of the continuum limit of modified loop models, and read, in the Kac-table notation of (\ref{confweights}),
\begin{eqnarray}
( \Delta,\bar{\Delta}) =( \Delta_{(1/2,-e_\phi)},\Delta_{(1/2,e_\phi)})\,.
\end{eqnarray}
The two special choices of $z$-factors give, after reshuffling the labels,  exponents $(\Delta_1,\bar{\Delta}_1)=(\Delta_{(2,1)},0)$ and $(\Delta_2,\bar{\Delta}_2)=(\Delta_{(0,1)},0)$. 

A more thorough study (to appear in the sequel of this paper) would show  that the second field disappears from the spectrum because the identity itself is degenerate\footnote{Meaning, its null descendent is zero indeed.} (and contributes as the field $\langle 1,1\rangle$) in the CFT.  The defect Hilbert space $\left({\cal H}_{O(n)}\right)_D$ \cite{CSLWY}  therefore has $(\Delta_{(2,1)},0)$ as its lowest excitation.%
\footnote{ This calculation holds in the dilute as well as in the dense case. In the latter case, however, our conventions where the Kac formula is parametrized by $\beta^2$ corresponds, in the conventions of \cite{BGJST}, to $(\Delta_{(1,2)},0)$, as found (using the Bethe ansatz) in this reference.}
It turns out, in fact, that the  whole $\left({\cal H}_{O(n)}\right)_D$  can   obtained by fusion with a degenerate field with those weights, so one has 
\begin{equation}
\left({\cal H}_{O(n)}\right)_D=\bigoplus_{s\in 2\mathbb{N}+1} \mathcal{R}_{\langle 2,1;s\rangle}\otimes [\,]\oplus \ldots \,, \label{defectHsp}
\end{equation}
where for simplicity we have refrained to get into the details of the whole field content.  The characters for the  corresponding representations indicated in  (\ref{defectHsp}) are 
\begin{equation}
 \chi_{\langle 2,1;s\rangle}(\tau) = \frac{e^{2\pi i\tau P^2_{(2,s)}} - e^{2\pi i\tau P^2_{(2,-s)}}}{\eta(\tau)}\times \frac{e^{2\pi i\bar\tau P^2_{(1,s)}} - e^{2\pi i\bar \tau P^2_{(1,-s)}}}{\eta(\bar\tau)}
 \end{equation}
 Of course, the defect $\bar{\DD}$ would give identical results with the chiral and antichiral sectors exchanged.

Notice that the geometry of a cylinder with the TDL running parallel to the imaginary time direction should be related with the one where the TDL acts as the operator  $\DD$ by modular transformation. This leads to the interpretation of,  for instance, $\langle \II|\DD|\II\rangle$ as a ratio of modular transformation matrix elements, something we will discuss in the sequel. The important point for now is that  our TDLs are  a close relative of Verlinde lines \cite{CSLWY}, although of course the underlying CFT is not rational. The same would hold for the higher-spin defects  (\ref{higherdef}), which are associated with a degenerate field with weights $(\Delta_{(2j+1,1)},0)$. Known fusion rules for these fields (which mimic (\ref{higherdef})) guarantee that the defect lines are not invertible, as claimed as the beginning of this paper.

\medskip

As an aside, we notice that the  field with weight $(\Delta_{(2,1)},0)$ is mutually local
with all the fields in the identity sector, in particular the diagonal primaries $\langle 1,s\rangle$, $s\in 2\mathbb{N}+1$. This extends to all ``higher'' disorder fields with conformal weights $(\Delta_{(r,1)},0)$ thanks to the identity
\begin{equation}
\Delta_{(r,1)}+\Delta_{(1,s)}-\Delta_{(r,s)}={(r-1)(s-1)\over 2} \,,
\end{equation}
which is an integer for $r$ integer, and $s\in 2\mathbb{N}+1$.\footnote{Similar properties were noticed in \cite{Delfino}, although the relationship with our work is not clear to us.}  
 
\section{Conclusion}

We conclude this paper with some general comments and possible generalizations.

\paragraph{Crossings versus self-attractions.} 

First, the reader might be confused by the fact that an operator that ``looks like the four-leg operator'' is often considered in the self-avoiding walk problem in the context of self-attracting chains. It is possible indeed to consider a general $O(n)$ loop model where neighboring segments are favored with some fugacity $\mu$ greater than one. At the level of the transfer matrix, this corresponds to multiplying by $\mu$ the last two terms in (\ref{Tmat}). As long as $\mu$ is not too large, the universality class remains the same (although the value of the critical coupling depends on $\mu$), but for a critical value  $\mu_c$ it becomes different and, in the case $n=0$, corresponds to ``polymers at the theta point''. It is important to understand that the operator $\tilde{\epsilon}$ that couples to the parameter $\mu$ is, despite the rough similarity of the geometry, {\sl not} the four-leg operator---precisely because it does not break the topological symmetry. On the lattice, the four-leg operator is represented instead by the last diagram in (\ref{diluteBrauerTM}), corresponding to the crossing of two lines or the operator denoted $\Pi$ in the $O(n)$ language. The operator $\tilde{\epsilon}$, by contrast,  is in the sector of the identity, and can be obtained by fusion of the energy operator with itself. In the dilute $O(n)$ model the energy operator corresponds to $\Phi_{\langle 1,3\rangle}$, a field which remains exactly degenerate at level $3$ (despite the theory being in general logarithmic): it follows that $\tilde{\epsilon}\propto \Phi_{\langle 1,5\rangle}$. Related aspects are discussed in \cite{Nahum}.

%

\paragraph{Classification of topological defects.}

As mentioned at the beginning of this paper, the loop  model admits  also invertible defects  \cite{CSLWY}  where the topological  lines are now associated with a group element of  $O(n)$. Such defects were already considered in  \cite{JRS22} and obtained  in the lattice formulation by modifying  the basic spin-spin coupling for bonds intersecting the defect line: $\vec{S}_x \cdot \vec{S}_y\mapsto \vec{S}_x \cdot g\vec{S}_y$ where $g \in O(n)$. For this to make sense, a convention is needed---say, after orienting the defect line,  that this applies for $x$  to the left of the line and $y$ to the right. In the plane, loops intersecting the defect line an even number of times do not get their weight modified, as a result of $O(n)$ symmetry. For instance, a loop intersecting the defect line twice will get a weight $\sum_{a,b} g_{ab}^2=\hbox{Tr }gg^t=n$, where $g^t$ denotes the transpose of $g$, and we used $gg^t=1$.

Denoting  such defects by $D(g)$, the opposite orientation of the defect line would define similarly $\overline{D}(g) := D(g^t)$.  In the plane, note that when such a defect line is inserted at a given point, loops surrounding this point get a modified weight $n\mapsto \hbox{Tr }g$. Hence it seems possible to think of the diagonal operators of \cite{Sylvain} as  operators sitting at the end of defect lines.%
\footnote{They are called ``defect operators'' standard terminology of \cite{CSLWY}, not to be confused with  the operator version of the TDL such as $\DD$.}
This aspect, how our non-invertible lines can be considered as Verlinde lines, and the more general question of classifying topological defects for the $O(n)$ CFT, will be discussed in the sequel.

We also notice that in the recent paper \cite{GSJNR23}, correlation functions in loop models were closely associated  with combinatorial maps. This suggests that there might yet be another, ``combinatorial''  type of defect in this problem.%
\footnote{S. Ribault, private communication}

\bigskip

Finally, we mention that it is an outstanding question whether one can make rigorous sense of TDLs
in terms of Deligne categories \cite{BR19}.  Here and in \cite{JRS22}, we have pursued the pragmatic
approach of making sense of computations with topological defect lines (be they invertible or non-invertible) for non-integer $n$ in terms of naive diagrammatic calculus and analytic continuation (in particular, when discussing RG flows), leaving for future work the question of whether this can be lifted to a more rigorous  setting. It is encouraging in this respect  that  the current in the $O(n)$ model---in some sense the infinitesimal form of a defect lines---has already been given a categorial sense in \cite{BR19}. We shall have  more to say about the $O(n)$ current in  other future  work.

\bigskip

\noindent {\bf Acknowledgments:} We thank L.\ Grans-Samuelsson, R.\ Nivesvivat and S.\ Ribault for related collaborations, many interesting discussions, and a careful reading of the manuscript.
We are also grateful to V.\ Gorbenko, S.\ Rychkov and B.\ Zan for remarks and correspondence.
This work was supported by the French Agence Nationale de la Recherche (ANR) under grant ANR-21-CE40-0003 (project CONFICA).

\appendix

\section{Allowing crossings at the dilute critical point}
\label{app:A}

In this appendix we present a numerical study of the stability of the $O(n)$ model towards the inclusion of some
amount of crossing. We first examine the global phase diagram, giving evidence that crossings are irrelevant
at the dilute critical point, whereas they are relevant in the dense phase and drive the system to the $\sigma$-model
fixed point shown in figure~\ref{Fig0}. Next we investigate in more details the dilute critical point, finding
notably that while the crossings lead to a decrease of the critical monomer fugacity $K_{\rm c}$, the universality
class, as measured by the central charge, is the same as in the model without crossings. And finally we study
the same issue from the point of view of critical exponents, finding that the crossings nevertheless lead to
more subtle and non-trivial algebraic effects that can be appreciated by comparing with the branching rules
from the (dilute) Brauer to the $\dJTL$ algebra.

\subsection{Model}

We define a {\em dilute Brauer model} on the square lattice with the following allowed vertex diagrams and weights:
$$
\label{diluteBrauerTM}
\begin{tikzpicture}[baseline=(current  bounding  box.center), scale = .5]
 \draw[dashed] (1,0) -- (1,2);
 \draw[dashed] (0,1) -- (2,1);
 \draw(1,-0.2) node[below] {$1$};
\end{tikzpicture}
\qquad
\begin{tikzpicture}[baseline=(current  bounding  box.center), scale = .5]
 \draw[dashed] (1,0) -- (1,2);
 \draw[dashed] (0,1) -- (2,1);
 \draw[blue,very thick] (1,0) -- (1,1) -- (2,1);
 \draw(1,-0.2) node[below] {$K$};
\end{tikzpicture}
\qquad
\begin{tikzpicture}[baseline=(current  bounding  box.center), scale = .5]
 \draw[dashed] (1,0) -- (1,2);
 \draw[dashed] (0,1) -- (2,1);
 \draw[blue,very thick] (2,1) -- (1,1) -- (1,2);
 \draw(1,-0.2) node[below] {$K$};
\end{tikzpicture}
\qquad
\begin{tikzpicture}[baseline=(current  bounding  box.center), scale = .5]
 \draw[dashed] (1,0) -- (1,2);
 \draw[dashed] (0,1) -- (2,1);
 \draw[blue,very thick] (1,2) -- (1,1) -- (0,1);
 \draw(1,-0.2) node[below] {$K$};
\end{tikzpicture}
\qquad
\begin{tikzpicture}[baseline=(current  bounding  box.center), scale = .5]
 \draw[dashed] (1,0) -- (1,2);
 \draw[dashed] (0,1) -- (2,1);
 \draw[blue,very thick] (0,1) -- (1,1) -- (1,0);
 \draw(1,-0.2) node[below] {$K$};
\end{tikzpicture}
\qquad
\begin{tikzpicture}[baseline=(current  bounding  box.center), scale = .5]
 \draw[dashed] (1,0) -- (1,2);
 \draw[blue,very thick] (0,1) -- (2,1);
 \draw(1,-0.2) node[below] {$K$};
\end{tikzpicture}
\qquad
\begin{tikzpicture}[baseline=(current  bounding  box.center), scale = .5]
 \draw[blue,very thick] (1,0) -- (1,2);
 \draw[dashed] (0,1) -- (2,1);
 \draw(1,-0.2) node[below] {$K$};
\end{tikzpicture}
\qquad
\begin{tikzpicture}[baseline=(current  bounding  box.center), scale = .5]
 \draw[blue,very thick] (1,0) -- (1,2);
 \draw[blue,very thick] (0,1) -- (2,1);
 \draw(1,-0.2) node[below] {$K^2 w$};
\end{tikzpicture}
$$
The meaning of the last diagram is that the two strands cross. The weights, distributed over
half-edges in the usual way, can be summarised by saying that there is a weight $K$ per monomer
and $w$ per crossing, and in addition a factor $n$ per loop.

Notice that even when $w=0$, this model does not quite simplify to that of (\ref{Fig6}). We have omitted
for simplicity the last two diagrams in (\ref{Fig6}) in which two distinct loop strands come into close contact
at at vertex. We have also rotated the vertices to make clear that we study the model on an axially oriented
square lattice (auxiliary-space geometry), wrapped onto a cylinder of circumference $L$ lattice sites.
The numerical technique is the exact diagonalization of the corresponding transfer matrix.

The dilute Brauer model with $w=0$ reduces in the $n \to 0$ limit
to a much-studied self-avoiding walk (SAW) model on the square
lattice. In this case the location of the critical point is known very precisely \cite{JSG16}
from the method of graph polynomials \cite{J14,J15},
\begin{equation}
 K_{\rm c} = 0.379052277755161 (5) \,,
\end{equation}
and exact enumeration studies have provided very compelling evidence \cite{JG99}
that the universality class is that of the dilute $O(n\to0)$ model.

For $n=1$ the loops are in bijection with the domain walls of an Ising model defined on the dual lattice.
Then $K$ is the weight of a piece of domain wall, and the range $w>1$ (resp.\ $0 \le w<1$) corresponds to a four-spin interaction that favours
(resp.\ penalises) the alternating arrangement $+-+-$ of spins arond a vertex. The case $w=1$ is just the usual
Ising model without such a four-spin interaction, and the critical point in then
\begin{equation}
 K_{\rm c} = \sqrt{2} - 1 \,.
\end{equation}

\subsection{Phase diagram}

For convenience we pick a generic value of $n$, far from the root of unity cases. We take $n = 1/\sqrt{2}$.
From the finite-size free energy per vertex $f(L)$ on cylinders of circumference $L$, $L-1$, $L-2$ we fit for
the bulk free energy $f(\infty)$, the effective central charge $c_{\rm eff}$ and the non-universal amplitude $A$ related
to $T \bar{T}$, via the formula
\begin{equation}
 f(L) = f(\infty) - \frac{\pi c_{\rm eff}}{6 L^2} + \frac{A}{L^4} \,.
\end{equation}
We can also omit the $1/L^4$ term and study fits based on only two sizes $L$, $L-1$.

\begin{figure}
\begin{centering}
 \includegraphics[width=0.45\textwidth]{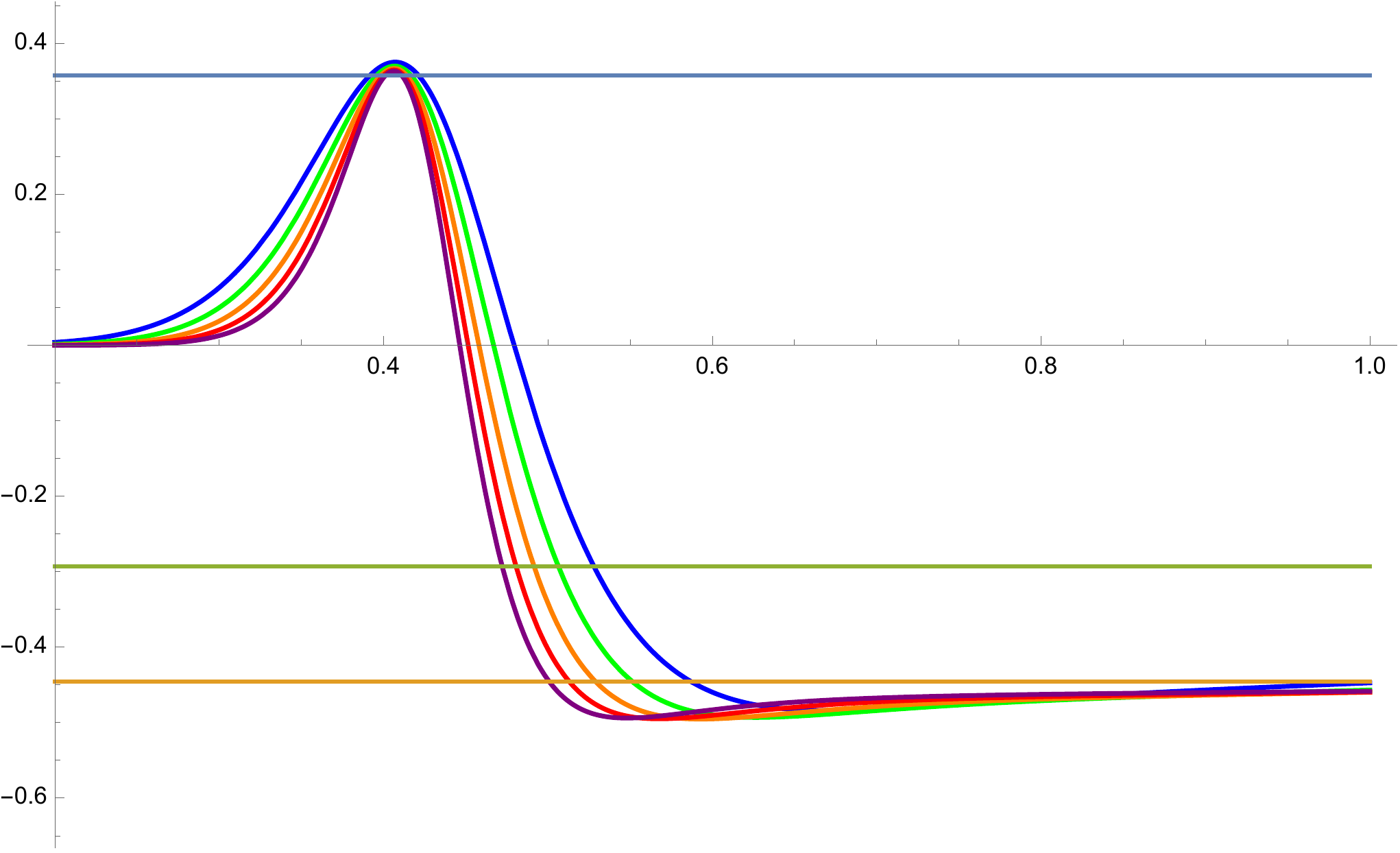}
 \includegraphics[width=0.45\textwidth]{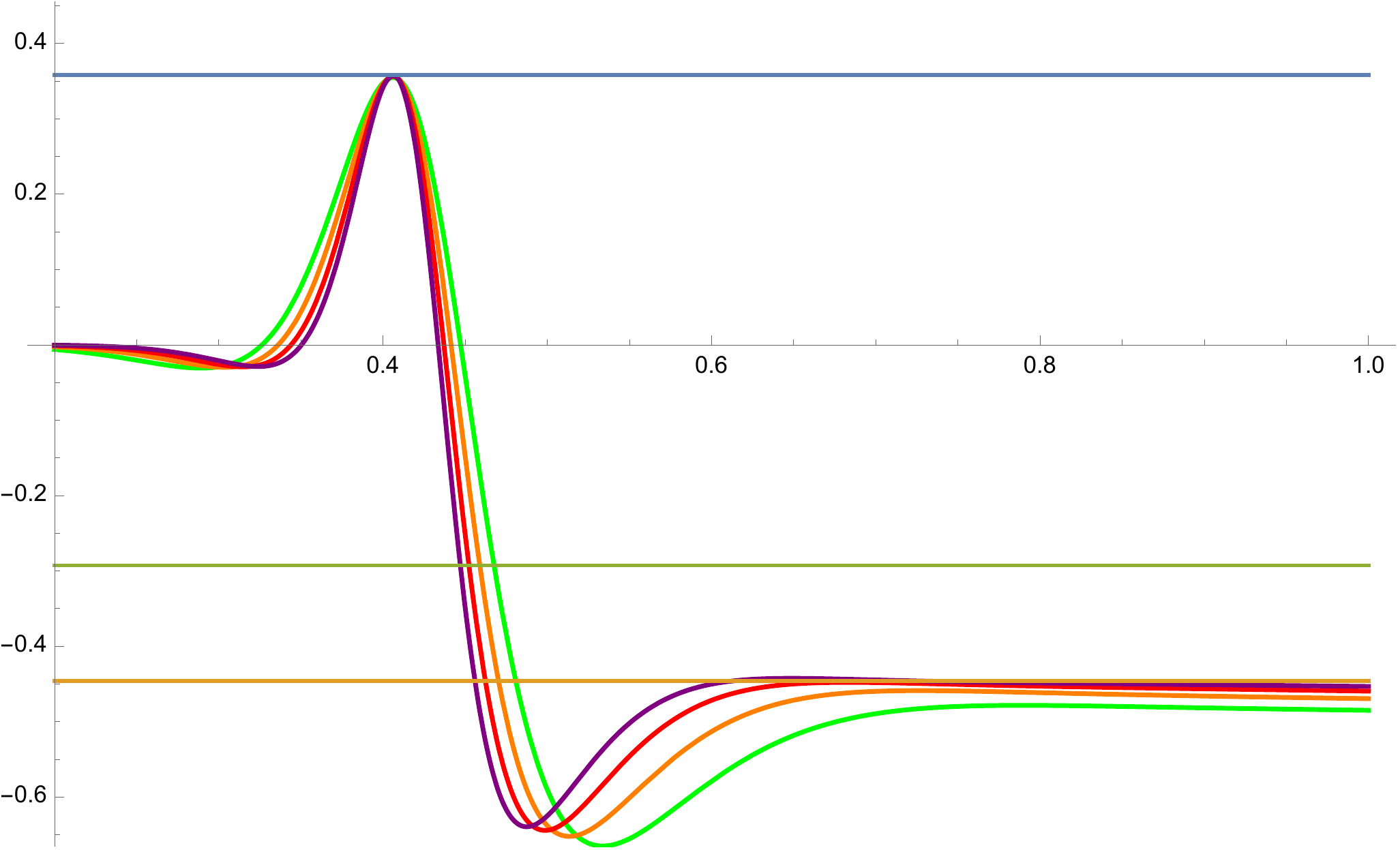} \\
 \includegraphics[width=0.45\textwidth]{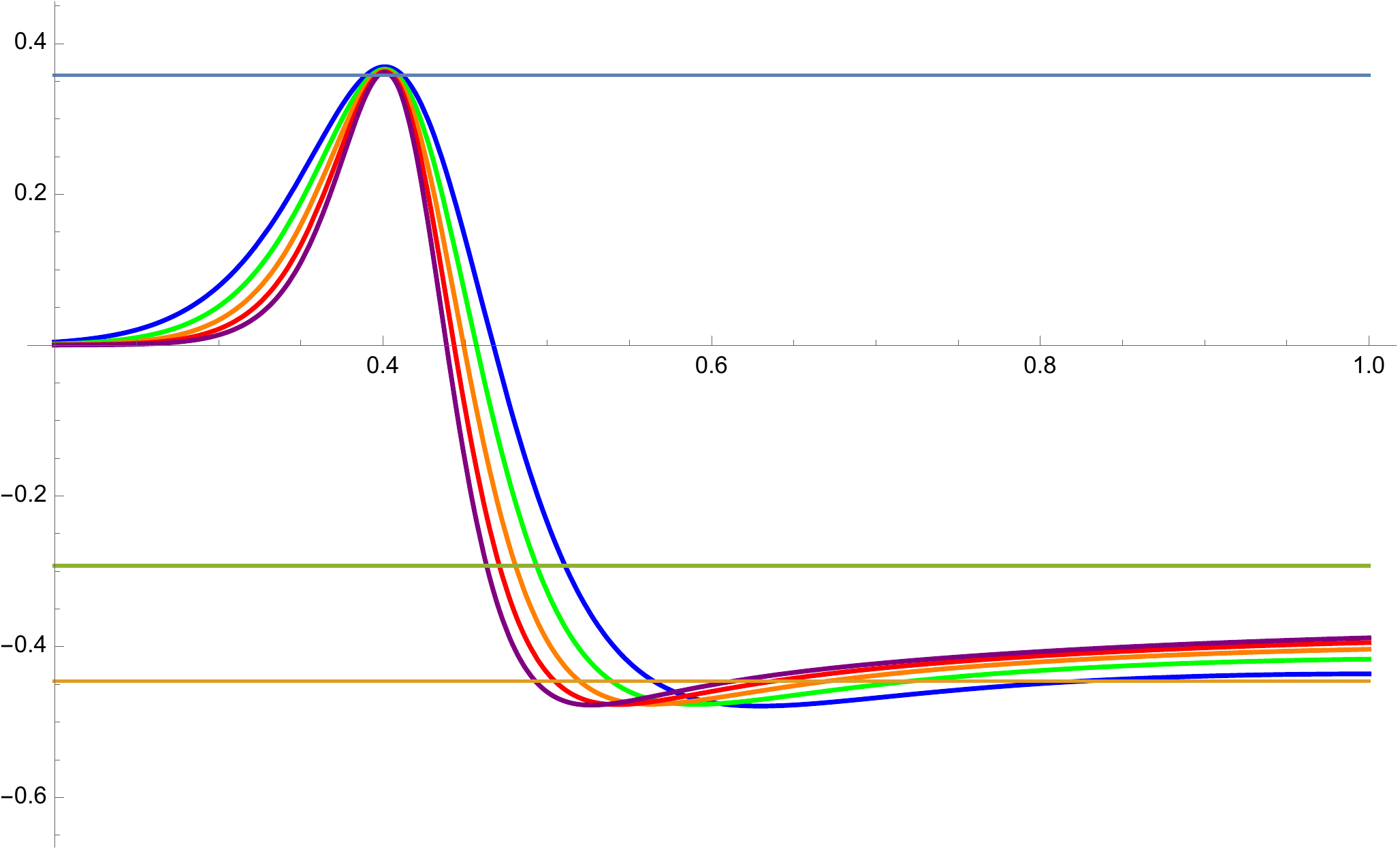}
 \includegraphics[width=0.45\textwidth]{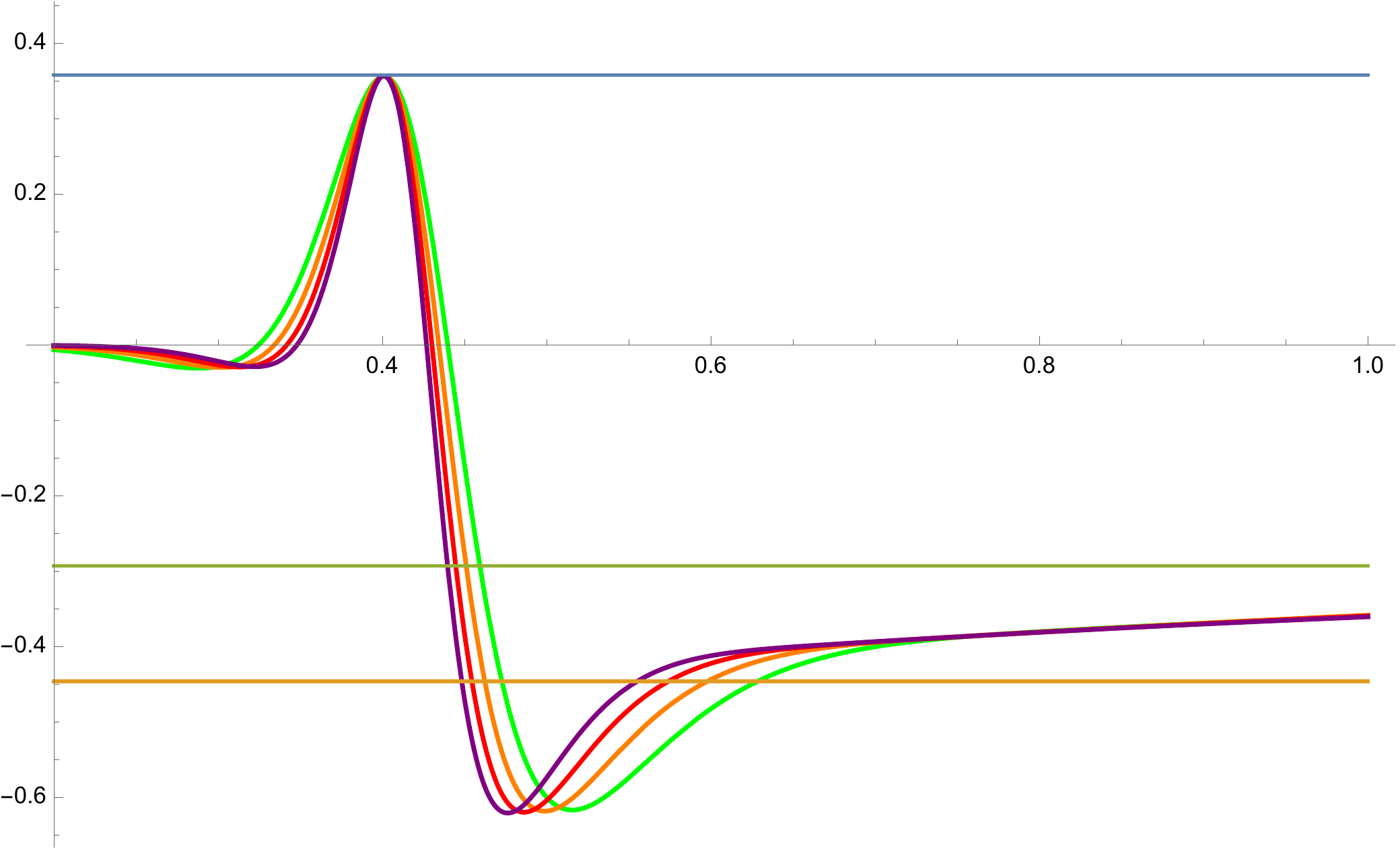} \\
 \includegraphics[width=0.45\textwidth]{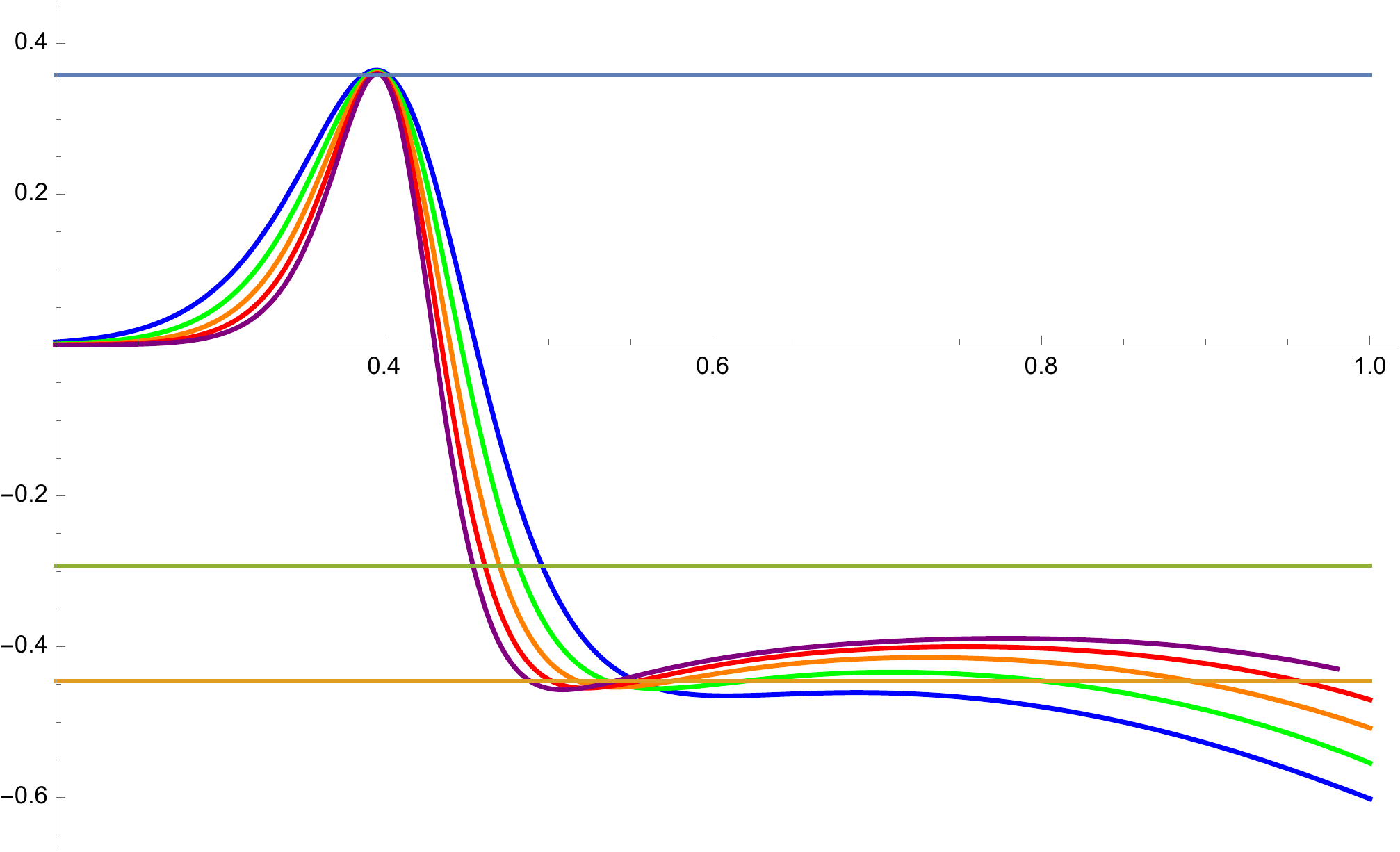}
 \includegraphics[width=0.45\textwidth]{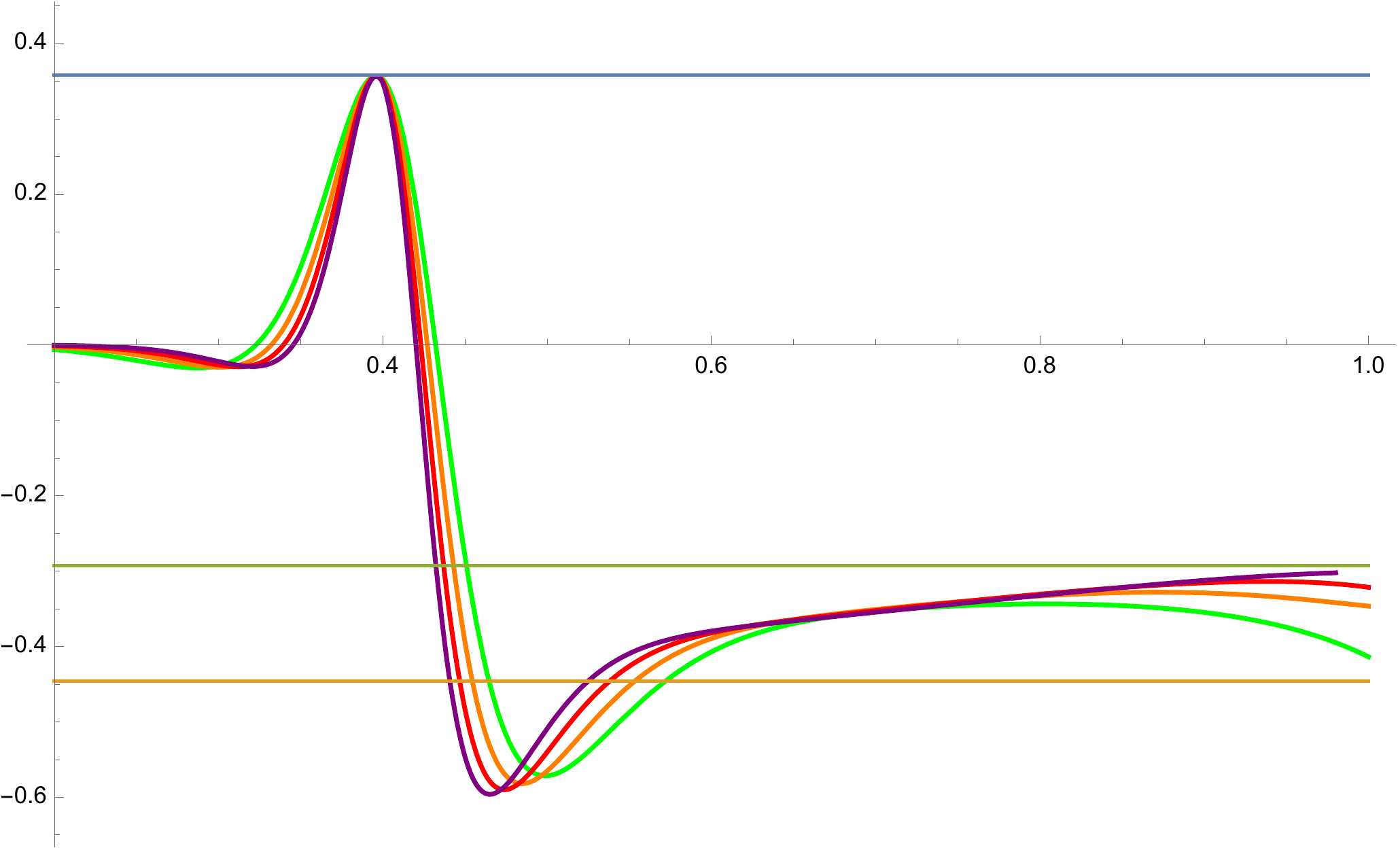} \\ 
\end{centering}
\caption{Effective central charge $c_{\rm eff}(K)$ of the dilute Brauer model. The left panels show two-point fits, while the right panels show three-point fits (with a $1/L^4$ term). The three rows of figures correspond to $w=0, 0.8, 1.6$. The horizontal lines indicate the analytically known $c$ in the dilute phase (upper line), dense phase (lower line) and Goldstone phase (middle line). All figures are for the generic loop weight $n=1/\sqrt{2}$. System sizes are $L=6,\ldots,11$, with the purple curves corresponding to the largest sizes.}
\label{fig:dense}
\end{figure}

We first study the global phase diagram and the flow from the dilute to the dense phase.
Effective central charges $c_{\rm eff}(K)$ are shown in Figure~\ref{fig:dense}.
The first row of figures corresponds to the case without crossings ($w=0$), in which the usual
flow from the dilute to the dense $O(n)$ fixed point is observed. The upper (resp.\ lower) horizontal line shows
for comparison the analytical value for the dilute (resp.\ dense) $O(n)$ model, namely $c = 0.357\,946\cdots$
(resp.\ $c = -0.445\,833\cdots$).

The next rows include crossings and show the cases $w=0.8, 1.6$.
They appear to be compatible with a flow to the Goldstone phase with central charge $c=n-1$ (shown
for comparison by the middle horizontal line). The convergence is slow, even for the three-point fits (right panels),
as was previously observed in the fully-dense case \cite{JReadS}.

\subsection{Universality class in the dilute case}

In the dilute case with $w \ge 0$, we observe a peak in $c_{\rm eff} = c_{\rm eff}(K)$, whose position gives a finite-size
estimate $K_{\rm c}(L)$ of the critical monomer weight and a finite-size estimate $c(L) = c_{\rm eff}(K_{\rm c}(L))$
of the central charge. Both of these estimates can be well extrapolated to the thermodynamic limit $L\to\infty$
by using a polynomial fit in the variable $1/L^2$, and error bars can be judged by comparing different orders of the fit.
Using data up to size $L=13$ we obtain the following values (the estimated error bars affect that last digit shown):
\begin{equation}
 \begin{tabular}{l|l|l}
 $w$ & $K_{\rm c}$ & $c$ \\ \hline
 0.0 & 0.406178 & 0.35784 \\
 0.1 & 0.405558 & 0.35787 \\
 0.2 & 0.404935 & 0.35790 \\
 0.4 & 0.403686 & 0.35794 \\
 0.8 & 0.401165 & 0.35798 \\
 1.6 & 0.396004 & 0.35802 \\
 3.2 & 0.385022 & 0.35789 \\
 \end{tabular}
\end{equation}

The value of $K_{\rm c}$ decreases with $w$, as it should. Moreover, $c$ agrees with the analytical
value for the dilute $O(n)$ model to four significant digits,
for any of the $w$ considered. Based on this evidence we conjecture that the dilute Brauer model
for any $0 \le n < 2$ and any finite $w \ge 0$ flows to the dilute $O(n)$ fixed point.

\subsection{Critical exponents}

We now turn to the study of critical exponents, still at the dilute critical point.
In the dilute Brauer model we impose the propagation of $2j$
throughlines (with $j \in \mathbb{Z}/2$) carrying an irrep $\lambda$ of the symmetric group ${\mathcal S}_{2j}$.
Experience teaches us that such a study is very sensitive to the precise determination of $K_{\rm c}$.
We therefore focus on the Ising case, $n=1$, for which the critical point $(w_{\rm c},K_{\rm c}) = (1,\sqrt{2}-1)$
is known analytically.

The methodology for obtaining precise estimates of the critical exponents has been explained in \cite[Appendix A.5]{JS18}.
The quality of the fits varies considerably with the number of available sizes. The preliminary results below are
based on data up to $L=11$ in all cases, $L=12$ in most cases, and $L=13$ for the $\lambda = [\,]$ sector.

The critical exponents for the Jones-Temperley-Lieb (JTL) standard modules $\Lambda_{(r,s)}$ to be compared with
are denoted
\begin{equation}
 x_{(r,s)} = \Delta_{(r,s)} + \Delta_{(r,-s)}
\end{equation}
with the Kac table for the Ising model being
\begin{equation}
 \Delta_{(r,s)} = \frac{(4r-3s)^2-1}{48} \,.
\end{equation}
We list here for reference the first few exponents for JTL primaries:
\begin{equation}
 \begin{tabular}{lll}
  $x_{(\frac12,0)} = \frac18 = 0.125$ \,, \\[2mm]
  $x_{(1,0)} = \frac58 = 0.625$ \,, & $x_{(1,1)} = 1$ \,, \\[2mm]
  $x_{(\frac32,0)} = \frac{35}{24} = 1.45833$ \,, \quad & $x_{(\frac32,\frac23)} = \frac{13}{8} = 1.625$ \,, \\[2mm]
  $x_{(2,0)} = \frac{21}{8} = 2.625$ \,, & $x_{(2,\frac12)} = \frac{87}{32} = 2.71875$ \,, \quad & $x_{(2,1)} = 3$ \,, \\[2mm]
  $x_{(\frac52,0)} = \frac{33}{8} = 4.125$ \,, & $x_{(\frac52,\frac25)} = \frac{837}{200} = 4.185$ \,. & \\
 \end{tabular}
\end{equation}

For the eigenvalues in the $\lambda = [\,]$ sector (no throughlines) we obtain the following results:
\begin{equation}
\label{expBJTL0}
 \begin{tabular}{r|l|l|l}
 No. & Mult. & $x$ (numerics) & $x$ (exact) \\ \hline
 1 & 1 & 0 & 0 \\
 2 & 1 & 0.99999999 & 1 \\
 3 & 4 & 2.0002 & 2 \\
 4 & 4 & 2.99995 & 3 \\
 5 & 1 & 2.66 & 2.625 \\
 6 & 1 & 3.0003 & 3 \\
 7 & 2 & 3.0006 & 3 \\
 8 & 4 & 3.97 & 4 \\
 9 & 2 & 3.64 & 3.625 \\
 10 & 4 & 4.0001 & 4 \\
 11 & 2 & ? & ? \\
 12 & 2 & ? & ? \\
 13 & 4 & 4.97 & 5 \\
 \end{tabular}
\end{equation}
where the columns denote the eigenvalue number, its multiplicity (valid for sufficiently large $L$, and in most cases
for all $L$), the corresponding numerically determined exponent (with a number a digits roughly reflecting the quality
of the fit), and finally the conjectured exact value (with question marks indicating uncertain cases).
These results seem consistent with the Brauer $\downarrow$ JTL branching rule \cite[eq.~(B.4a)]{JRS22}
\begin{equation}
B_{[\,]} \downarrow \overline{W}_0 + W_{(2,0)} + \cdots \,.
\end{equation}
Notice that we also observe a descendent state in the $W_{2,0}$ module.
Although multiplicities would have to be explained, the results (\ref{expBJTL0}) confirm that
the inclusion of crossings at the dilute critical point preserves not only the central charge but also
the critical exponents, up to the expected reshuffling of sectors implied by the branching rules.

For eigenvalues in the $\lambda = [1]$ sector we obtain:
\begin{equation}
 \begin{tabular}{r|l|l|l}
 No. & Mult. & $x$ (numerics) & $x$ (exact) \\ \hline
 1 & 1 & 0.124999 & 0.125 \\
 2 & 2 & 1.12497 & 1.125 \\
 3 & 2 & 1.633 & 1.625 \\
 4 & 2 & 2.124 & 2.125 \\
 5 & 2 & 2.15 & 2.125 \\
 6 & 1 & 2.124998 & 2.125 \\
 7 & 2 & 2.6 & 2.625 \\
 8 & 1 & 2.62 & 2.625 \\
 9 & 2 & 3.11 & 3.125 \\
10 & 2 & 2.7 ? & 2.625 \\
11 & 1 & 2.64 & 2.625 \\
 \end{tabular}
\end{equation}
This already appears more surprising. The branching rule is here \cite[eq.~(B.4b)]{JRS22}
\begin{equation}
 B_{[1]} \downarrow W_{(\frac12,0)} + W_{(\frac52,0)} + \cdots \,.
\end{equation}
But we see here with a good precision an exponent $x = 1.625$, which must correspond to a primary
(since there is no $x = 0.625$), and hence it can only be $x_{(\frac32,\frac23)}$, although the
representation $W_{(\frac32,\frac23)}$ does not appear on the right-hand side of the branching rule.
We also observe several descendents of that primary.

It should be noticed that if we had diagonalised a transfer matrix for a loop model {\em without} crossings
(i.e., an element of the $\dJTL$ algebra) within a given standard module of the
dilute Brauer algebra, we would {\em by definition} obtain the critical exponents of the dilute model
up to the reshuffling implied by the branching rules. As a matter of fact, this is precisely how the
branching rules were obtained in \cite[Appendix B]{JRS22}, albeit in the fully-dense incarnations of
both models (but the results should carry over unchanged by Morita equivalence). However, this is
not quite what is done here, since we use a transfer matrix in a loop model where crossings are now allowed.
Even if those crossings are not expected to change the universality class,
they could---and as we have just observed, do---change the identification of critical exponents within
each representation $\lambda$.

We leave for future work the analytical understanding of exactly how critical exponents are reshuffled within
this scenario, and here content ourselves with reporting the numerical results for a few more sectors.
For eigenvalues in the $\lambda = [2]$ sector we obtain:
\begin{equation}
 \begin{tabular}{r|l|l|l}
 No. & Mult. & $x$ (numerics) & $x$ (exact) \\ \hline
 1 & 1 & 0.622 & 0.625 \\
 2 & 2 & 1.620 & 1.625 \\
 3 & 2 & 2.12 & ? \\
 4 & 2 & 2.63 & 2.625 \\
 5 & 1 & 2.61 & 2.625 \\
 6 & 2 & 2.6 & 2.625 \\
 7 & 2 & 3.2 ? & ? \\
 \end{tabular}
\end{equation}
The branching rule is now $B_{[2]} \downarrow W_{(1,0)} + W_{(2,0)} + W_{(2,1)} + \cdots$. The third exponent in the
table is surprising and it is not yet clear what it corresponds to. The seventh exponent is possibly its descendent,
but the quality of the fit is not very convincing.
We would need more eigenvalues to confirm the presence of $W_{2,1}$.

For eigenvalues in the $\lambda = [11]$ sector we obtain:
\begin{equation}
 \begin{tabular}{r|l|l|l}
 No. & Mult. & $x$ (numerics) & $x$ (exact) \\ \hline
 1 & 2 & 1.002 & 1 \\
 2 & 2 & 1.99 & 2 \\
 3 & 2 & 3.08 & 3 ? \\
 4 & 2 ? & 4.06 & 4 \\
 \end{tabular}
\end{equation}
There are many complex eigenvalues in this sector which we have omitted from the fits. These do {\em not} appear
to move deeper into the spectrum upon increasing $L$. In particular, the first complex eigenvalues (part of a complex
conjugate doublet) is number $8, 9, 7, 5$ (counting multiplicities in the numbering) in the spectrum sorted by largest real part for the
sizes $L=8, 9, 10, 11$. The branching rule is here $B_{[1,1]} \downarrow W_{(1,1)} + W_{(2,\pm \frac12)} + \cdots$. More
eigenvalues would be needed to confirm the presence of $W_{(2,\frac12)}$.

For eigenvalues in the $\lambda = [3]$ sector we obtain:
\begin{equation}
 \begin{tabular}{r|l|l|l}
 No. & Mult. & $x$ (numerics) & $x$ (exact) \\ \hline
 1 & 1 & 1.449 & 1.45833 \\
 2 & 2 & 2.447 & 2.45833 \\
 3 & 2 & 2.95 & ? \\
 4 & 2 & 3.44 & 3.45833 \\
 5 & 2 & 3.46 & 3.45833 \\
 6 & 1 ? & 3.42 & 3.45833 \\
 \end{tabular}
\end{equation}
The branching rule is $B_{[3]} \downarrow W_{(\frac32,0)} + W_{(\frac52,0)} + \cdots$. The third exponent in the table
is surprising. The quality of the corresponding fit is reasonable, but not excellent. More
eigenvalues would be needed to confirm the presence of $W_{(\frac52,0)}$.

For eigenvalues in the $\lambda = [21]$ sector we obtain:
\begin{equation}
 \begin{tabular}{r|l|l|l}
 No. & Mult. & $x$ (numerics) & $x$ (exact) \\ \hline
 1 & 2 & 1.633 & 1.625 \\
 2 & 2 & 2.15 & ? \\
 3 & 2 & 2.59 & 2.625 \\
 4 & 1 & 2.638 & 2.625 \\
 5 & 1 & ? & ?  \\
 6 & 2 ? & ? & ? \\
 \end{tabular}
\end{equation}
The branching rule is $B_{[2,1]} \downarrow 2W_{(\frac32,\pm \frac23)} + 2W_{(\frac52,0)} + 4W_{(\frac52,\pm \frac25)} + \cdots$.
The second exponent is surprising, but the fit is well behaved.
The last two fits are uncommonly capricious, even with $L=12$ data being available, so I refrain from reporting any value of the corresponding exponents.

Finally, for eigenvalues in the $\lambda = [111]$ sector we obtain:
\begin{equation}
 \begin{tabular}{r|l|l|l}
 No. & Mult. & $x$ (numerics) & $x$ (exact) \\ \hline
 1 & 1 & 1.48 & 1.45833 \\
 2 & 2 & 2.456 & 2.45833 \\
 3 & 2 & 3.0 ? & ? \\
 4 & 2 & ? & ? \\
 5 & 2 & 3.4 ? & 3.45833 ? \\
 6 & 2 & 3.9 ? & ? \\ 
 \end{tabular}
\end{equation}
A very encouraging fact is that this resembles the results for $\lambda = [3]$. Indeed, the representations
$B_{[3]}$ and $B_{[111]}$ have the same branching rules with respect to JTL. It should be stressed that
the finite-size eigenvalues are completely different for the two cases when (as here) $w > 0$. In particular,
the first eigenvalue in the $[3]$ sector gives rise to an exponent that converges to $x_{(\frac32,0)}$ from below,
whereas the first exponent from the $[111]$ sector converges to that same value from above.

\section{The non-invertible symmetry with open boundary conditions}

The study of the $O(n)$ model on a strip (i.e.\ with open boundary conditions) reveals a pattern of degeneracies which is similar to---but different from---the pattern on the cylinder. In this case, a genuine symmetry has been identified to explain these degeneracies, and involves  mixing of $O(n)$ representations under an algebra  equivalent in a certain (Morita) sense to $U_qsl(2)$: in particular, for this symmetry, tensor products can be defined, and the multiplicities be written as $q$-dimensions. 

\begin{figure}
\centering
\begin{tikzpicture}[scale=0.5]
 \draw[thick,red] (3.8,0) arc(0:45:1.0) -- (1,3) arc(225:135:1.0) -- (1.5,4.87);
 \draw[thick,red] (1.7,5.0) -- (2.9,6.0);
 \draw[thick,red] (3.1,6.15) -- (3.5,6.5) arc(-45:0:1.0);
 \draw[thick,red,dashed] (3.8,7.2) arc (180:0:0.3 and 0.5) -- (4.4,0) arc(0:-180:0.3 and 0.5);
 \draw (4.4,3.6) node[right] {\footnotesize Tr};
 \draw[thick,blue] (1.6,0) -- (1.6,2.3);
 \draw[thick,blue] (1.6,2.6) -- (1.6,7.2);
 \draw[thick,blue] (3,0) -- (3,1.05);
 \draw[thick,blue] (3,1.3) -- (3,7.2);
\end{tikzpicture} 
\caption{The $\dd$-operator in the open case.}
\label{Fig11}
\end{figure}
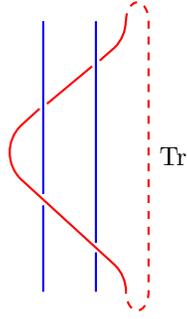

It is tempting nevertheless to see whether the consideration of a possible topological symmetry in this case might lead to further insight. 
In the open case, we cannot have a loop going around the system since the algebra does not contain the shift operator $\tau$. Instead, what we can have is a loop going over the lines, bouncing on the boundary, then going under and bouncing again. This is in fact described in detail in \cite{CGS22}, where it is shown that this defect  (denoted $\dd$ here)  is diagonal on every $\TL$  module  $W_r$ \footnote{Recall that these are modules with $2r$ throughlines.} with eigenvalue
\begin{equation}
\langle \II_{r}|\dd|\II_{r}\rangle=q^{2r+1}+q^{-2r-1} \,,
\end{equation}
that is, $\dd$ acts in fact as the Casimir of $U_qsl(2)$. We can illustrate this using  $O(n)$ variables: for instance on two sites,
\begin{equation}
\dd|ab\rangle=\sum_c I\otimes I\otimes \langle c| g_2g_1^2g_2|abc\rangle \,,
\end{equation}
as illustrated in figure \ref{Fig11}, where we have rewritten (\ref{Fig1}) as
\begin{equation}
g_i=(-q)^{1/2}+(-q)^{-1/2}e_i \,.
\end{equation}
A short calculation gives
\begin{equation}
\dd|ab\rangle=(q^3+q^{-3})\left(|ab\rangle-{\delta_{ab}\over n}\sum_c |cc\rangle\right)+n {\delta_{ab}\over n}\sum_c |cc\rangle
\end{equation}
with $n=q+q^{-1}$. We see that $\dd$ is equal to the $U_qsl(2)$ Casimir, with eigenvalue $q^3+q^{-3}$ on $[2]$ and $q+q^{-1}$ on $[\,]$.

In general we can write schematically
\begin{equation}
\dd=\sum_r (q^{2r+1}+q^{-2r-1})P_r \,,
\end{equation}
where $P_r$ is the Jones-Wenzl projector onto the sector with $2r$ non-contractible lines. In this case, therefore, the potential ``topological symmetry'' simply reveals the underlying quantum-group structure of the problem. Since in the periodic case there is no such known structure, it is likely that the topological defect, in a sense, replaces it.

\section{TDLs with crossings: the (dilute)  Brauer algebra case}

As commented earlier, the defect operator $\DD$ is built out of the  solution of the spectral-parameter independent Yang-Baxter equation in the Temperley-Lieb algebra. It is tempting to ask what would happen if we considered the same question in the case of the Brauer algebra. 

Consider therefore the  $O(n)$ integrable $\check{R}$ matrix acting in $[1]^{\otimes 2}$:
\begin{equation}
\check{R}(u)=P_{[\,]}-{u-i\pi\over u+i\pi}P_{[11]}+{u-i\pi\over u+i\pi}{(\n-2)u-i\pi\over (\n-2)u+i\pi}P_{[2]} \,.
\end{equation}
Solutions of the spectral-parameter independent Yang-Baxter equation are obtained by considering the limit $u\to\infty$, where  we obtain
\begin{equation}
\check{R}(u=\infty)=P_{[\,]}-P_{[11]}+P_{[2]}=\Pi \,.
\end{equation}
This is, in fact, the simple permutation operator, represented by two lines crossing. Hence, for Brauer, our construction is not relevant: the ``topological loop'' is just a loop crossing all the other lines, and it can be eliminated with a factor   $n$, due to the general rules of the Brauer algebra. 

The knowledgeable reader might also point out that potential topological defect operators can be obtained by considering the center of the  algebra underlying the transfer matrix. This object for the Brauer algebra  has of course been considered in the mathematics literature, and gives rise to the consideration of Jucys-Murphy elements (see e.g.\ \cite{IMR}). In the end, the consideration of the center allows one to determine separately every irreducible $O(n)$ representation that appears in the tensor product $[1]^{\otimes N}$. Like in the open case, it thus reveals in a certain sense the underlying $O(n)$ structure potentially hidden beyond the loop model, but does not provide additional information. 

In contrast, when crossings are not allowed, we get an algebra (the $\uJTL$  or $\dJTL$ algebra)   which is smaller than Brauer, and thus  the centralizer has to be bigger---leading to the multiplicities observed in the $O(n)$ model partition functions.


\begin{thebibliography}{99}
\bibitem{GZ} V.\ Gorbenko and B.\ Zan, ``Two-dimensional $O(n)$ models and logarithmic CFTs'', J.\ High Energ.\ Phys.\ {\bf 2020}, 99 (2020); {\tt arXiv:2005.07708}.
\bibitem{JRS22} J.L.\ Jacobsen, S.\ Ribault and H.\ Saleur, ``Spaces of states of the two-dimensional $O(n)$ and Potts models'',
 SciPost Phys., in press (2023);  {\tt arXiv:2208.14298}.
\bibitem{Nienhuis} B.\ Nienhuis, ``Exact critical point and critical exponents of $O(n)$ 
 models in two dimensions'', Phys.\ Rev.\ Lett.\ {\bf 49}, 1062 (1982).
\bibitem{DuplantierSaleur} B.\ Duplantier and H.\ Saleur, ``Exact critical properties of two-dimensional dense self-avoiding walks'',  Nucl.\ Phys.\ B {\bf 290}, 291 (1987).
\bibitem{JReadS}  J.L.\ Jacobsen, N.\ Read and H.\ Saleur,
 ``Dense loops, supersymmetry, and Goldstone phases in two dimensions'',
 Phys.\ Rev.\ Lett.\ {\bf 90}, 090601 (2003); {\tt arXiv:cond-mat/0205033}.
\bibitem{NSSO13} A.\ Nahum, P.\ Serna, A.M.\ Somoza and M.\ Ortu\~{n}o,
 ``Loop models with crossings'', Phys.\ Rev.\ B {\bf 87}, 184204 (2013); {\tt arXiv:1303.2342}.
\bibitem{Granet19} E.\ Granet, J.L.\ Jacobsen and H.\ Saleur,
 ``Spontaneous symmetry breaking in 2D supersphere sigma models and applications to intersecting loop soups'',
  J.\ Phys.\ A:\ Math.\ Theor.\ {\bf 52}, 345001 (2019); {\tt arXiv:1810.07807}.
\bibitem{CSLWY} C.M.\ Chang, Y.H.\ Lin, S.H.\ Shao, Y.\ Wang and X.\ Yin,
 ``Topological  defect lines and renormalization group flows in two dimensions'',
 J.\ High Energ.\ Phys.\ {\bf 2019}, 26 (2019); {\tt arXiv:1802.04445}.
 \bibitem{PetkovaZuber} 
 V.B.\ Petkova and J.-B.\ Zuber, ``Generalised twisted partition functions'', Phys.\ Lett.\ B {\bf 504}, 157 (2001); {\tt arXiv:hep-th/0011021}.
 \bibitem{Juergen} 
J.\  Fr\"ohlich, J.\  Fuchs, I.\  Runkel and Ch.\ Schweigert, ``Duality and defects in rational
conformal field theory,'' Nucl.\ Phys.\ B {\bf 763}, 354 (2007); {\tt arXiv:hep-th/0607247}.
\bibitem{AMF} D.\ Aasen, R.S.K.\ Mong and P. Fendley, ``Topological defects on the lattice I: The Ising model'',
 J.\ Phys.\ A:\ Math.\ Theor.\ {\bf 49}, 354001 (2016);  {\tt arXiv:1601.07185}.
\bibitem{BKW76} R.J.\ Baxter, S.B.\ Kelland and F.Y.\ Wu, ``Equivalence of Potts model or Whitney polynomial with an ice-type model'', J.\ Phys.\ A:\ Math.\ Gen.\ {\bf 9}, 397 (1976).
\bibitem{KGN96} J.\ Kondev, J.\ de Gier and B.\ Nienhuis,  ``Operator spectrum and exact exponents of the fully packed loop model'', J.\ Phys.\ A:\ Math.\ Gen.\ {\bf 29}, 6489 (1996); {\tt arXiv:cond-mat/9603170}.
\bibitem{JK98} J.L.\ Jacobsen and J.\ Kondev, ``Field theory of compact polymers on the square lattice'', Nucl.\ Phys.\ B {\bf 532}, 635--688 (1998); {\tt arXiv:cond-mat/9804048}.
\bibitem{BR19} D.J.\ Binder and S.\ Rychkov,
``Deligne categories in lattice models and quantum field theory'',
 J.\ High Energ.\ Phys.\ {\bf 2020}, 117 (2020); {\tt arXiv:1911.07895}.
\bibitem{BGJST} J.\ Bellet\^ete, A.M.\ Gainutdinov, J.L.\ Jacobsen, H.\ Saleur and T.S.\ Tavares, ``Topological defects in lattice models and affine Temperley-Lieb algebra'', 
 Comm.\ Math.\ Phys., in press (2023); {\tt arXiv:1811.0255}.
 \bibitem{Delfino} G.\ Delfino, ``Fields, particles and universality in two dimensions'', Ann.\ Phys.\ {\bf 333}, 1 (2013); {\tt arXiv:1502.05538}.
\bibitem{IMR} N.\ Iorgov, A.I.\ Molev and E.\ Ragoucy, ``Casimir elements from the Brauer-Schur-Weyl duality'', J.\ Algebra {\bf 387}, 144 (2013).
\bibitem{CGS22} D.\ Chernyak, A.\ Gainutdinov and H.\ Saleur, ``$U_q sl_2$-invariant non-compact boundary conditions for the XXZ spin chain'', J.\ High Energ.\ Phys.\ {\bf 2022}, 16 (2022); {\tt arXiv:2207.12772}.
\bibitem{Trebstetal} S.\ Trebst, E.\ Ardonne, A.\ Feiguin, D.A.\ Huse, A.W.W.\ Ludwig and M.\ Troyer, ``Collective states of interacting Fibonacci anyons'', 
Phys.\ Rev.\ Lett.\ {\bf 101}, 050401 (2008); erratum ibid.\ {\bf 101}, 149901 (2008); {\tt arXiv:0801.4602}.
\bibitem{BuicanGromov} M.\ Buican and A.\ Gromov, ``Anyonic chains, topological defects and conformal field theory'', Comm.\ Math.\ Phys.\ {\bf 356}, 1017 (2017); {\tt arXiv:1701.02800}.
\bibitem{Sylvain} S.\ Ribault, ``Diagonal fields in critical loop models'', SciPost Phys.\ Core {\bf 6}, 020 (2023); {\tt arXiv:2209.09706}.
\bibitem{OP} A.L.\ Owczarek and T.\ Prellberg,
``Collapse transition of self-avoiding trails on the square lattice'',
 Physica A {\bf 373}, 433--438 (2007); {\tt arXiv:cond-mat/0603405}.
\bibitem{JSG16} J.L.\ Jacobsen, C.R.\ Scullard and A.J.\ Guttmann, ``On the growth constant for square-lattice self-avoiding walks'', J.\ Phys.\ A:\ Math.\ Theor.\ {\bf 49}, 494004 (2016); {\tt arXiv:1607.02984}.
\bibitem{J14} J.L.\ Jacobsen, {\em High-precision percolation thresholds and Potts-model critical manifolds from graph polynomials}, J.\ Phys.\ A:\ Math.\ Theor.\ {\bf 47}, 135001 (2014); {\tt arXiv:1401.7847}.
\bibitem{J15} J.L.\ Jacobsen, {\em Critical points of Potts and $O(N)$ models from eigenvalue identities in periodic Temperley-Lieb algebras}, J.\ Phys.\ A:\ Math.\ Theor.\ {\bf 48}, 454003 (2015); {\tt arXiv:1507.03027}.
\bibitem{JG99} I.\ Jensen and A.J.\ Guttmann, ``Self-avoiding polygons on the square lattice'', J.\ Phys.\ A:\ Math.\ Gen.\ {\bf 32}, 4867 (1999); {\tt arXiv:cond-mat/9905291}.
\bibitem{JS18} J.L.\ Jacobsen and H.\ Saleur, ``Bootstrap approach to geometrical four-point functions in the two-dimensional critical $Q$-state Potts model: A study of the $s$-channel spectra'', J.\ High Energ.\ Phys.\ {\bf 2019}, 84 (2019); {\tt arXiv:1809.02191}.
\bibitem{GSJNR23} L.\ Grans-Samuelsson, J.L.\ Jacobsen, R.\ Nivesvivat, S.\ Ribault and  H.\ Saleur, ``From combinatorial maps to correlation functions in loop models'', {\tt arXiv:2302.08168}.
\bibitem{Nahum} A.\ Nahum, ``The generic critical behaviour for 2D polymer collapse'', Phys.\ Rev.\ E {\bf 93}, 052502 (2016); {\tt arXiv:1510.09223}.
\end{thebibliography}
\end{document}